\documentclass[12pt]{article}

\advance\topmargin by -1.6cm

\newenvironment{eq}[1]
{\[\begin{array}{#1}}{\end{array}\]}

\let\rvec=\vec        

\def\plint {\hskip-2mm{^\pl\hskip-3mm\int}}

 \def\({\Bigl(}
\def\){\Bigr)}   
 \def\|{\Big|}
\def\then{\Rightarrow}  
 \def\o{\circ}
\def\m{\bullet}    
\def\x{\times}
   
\def\ox{\otimes}

\def\pl{{~\oplus~}}

\def\SUM{\displaystyle \sum}

\def\mid{\big\bracevert}

\def\sub{\subseteq}
\def\subnoteq{\subset}
\def\sup{\supseteq}
\def\supnoteq{\supset}
\def\and{\wedge}

\def\rin{{\,\in\kern-.42em\in}}

\def\tr{{\,{\rm tr }\,}}
\def\det{\,{\rm det }\,}
\def\id{\,{\rm id}}

\def\Int{\,{\rm Int}\,}
\def\Ad{\,{\rm Ad}\,}
\def\ad{{\,{\rm ad}\,}}
\def\centr{\,{\rm centr}\,}

\def\card{\ro{\,card\,}}

\def\sx{~\rvec\x~\!}

\def\irrep{{{\bf irrep\,}}}

\def\A{{\,{\rm A\kern-.55emA}}}
\def\B{{\,{\rm I\kern-.2emB}}}
\def\C{{\,{\rm I\kern-.55emC}}}
\def\E{{\,{\rm I\kern-.2emE}}}
\def\G{{\,{\rm I\kern-.55emG}}}
\def\H{{{\rm I\kern-.2emH}}}
\def\I{{\,{\rm I\kern-.2emI}}}
\def\K{{\,{\rm I\kern-.2emK}}}
\def\L{{\,{\rm I\kern-.2emL}}}
\def\M{{\,{\rm I\kern-.16emM}}}
\def\N{{\,{\rm I\kern-.16emN}}}
\def\Q{{\,{\rm I\kern-.5emQ}}}
\def\R{{{\rm I\kern-.2emR}}}
\def\S{{\,{\rm I\kern-.42emS}}}
\def\T{{\,{\rm I\kern-.37emT}}}
\def\UU{{\,{\rm I\kern-.51emU}}}
\def\Z{{\,{\rm Z\kern-.32emZ}}}

\def\p{\partial}

\def\al{\alpha}  \def\be{\beta} \def\ga{\gamma}
\def\de{\delta}  \def\ep{\epsilon}  
\def\th{\theta}   \def\vth{\vartheta} \def\io{\iota}
   \def\la{\lambda}   \def\si{\sigma}
\def\De{\Delta}   \def\om{\omega} \def\Om{\Omega}
\def\phi{\varphi} 
 \def\Ga{\Gamma}  
    \def\La{\Lambda}

\def\vec#1{\underline{\bf vec}_{#1}}

\def\GL{{\bf GL}}  
\def\SL{{\bf SL}}
\def\U{{\bf U}} 
 
\def\O{{\bf O}}   
\def\SU{{\bf SU}} 
 
\def\SO{{\bf SO}}
 \def\Sp{{\bf Sp}} 
 
 \def\D{{\bl D}}
\def\AL{{\bf AL}}

\def\rstate#1{|#1\rangle}
\def\lstate#1{\langle#1|}

\def\brack#1{\lbrack#1\rbrack}

\def\ty#1{{\tt #1}}
\def\ro#1{{\rm #1}}
\def\bl#1{{\bf {#1}}}
\def\cl#1{{\cal #1}}
\def\ul#1{\underline{#1}}

\def\ol#1{\overline{#1}}

\def\lvec#1{\stackrel{\leftarrow}{#1}}

\def\dprod#1#2{\langle#1,#2\rangle}
\def\sprod#1#2{\langle#1|#2\rangle}

\def\map{\longrightarrow}

\def\lrmap{\leftrightarrow}

\def\mape{\longmapsto}

\begin{document}

\begin{titlepage} 
$~$
\vskip5mm
\hfill MPP-2004-130
\vskip25mm

\centerline{\bf THE BASIC PHYSICAL LIE OPERATIONS}
\vskip15mm
\centerline{
Heinrich Saller\footnote{\scriptsize hns@mppmu.mpg.de} }
\centerline{Max-Planck-Institut f\"ur Physik}
\centerline{Werner-Heisenberg-Institut}
\centerline{M\"unchen, Germany}

\vskip25mm

\centerline{\bf Abstract}
\vskip5mm

Quantum theory can be formulated as a theory of operations,
more specific, of complex represented operations from real Lie groups.  
Hilbert space eigenvectors of acting Lie operations
are used as  states or  particles.
The simplest simple Lie groups have three dimensions.
These groups together with their contractions
and their subgroups contain -  in the simplest form - 
all physically important basic operations which come as  translations
for causal time, for space and for spacetime, as
rotations, Lorentz transformations and as Euclidean and Poincar\'e transformations 
with scattering and particle states and also - via the Heisenberg group - 
as the operational 
structure of nonrelativistic quantum mechanics.
The classification of all those groups and their contractions  is given 
together with their
Hilbert spaces, constituted by energy-mo\-men\-tum functions.
The group re\-pre\-sen\-ta\-tion matrix elements 
can be written in the form of residues of energy-mo\-men\-tum poles 
- simple poles
for abelian translations, e.g. in Feynman propagators,
 and dipoles for simple group operations, e.g.
 in the Schr\"odinger wave functions for the nonrelativistic hydrogen atom.

\end{titlepage}

\newpage 
\setcounter{page}{0}
{\small\tableofcontents}
\newpage
\setcounter{page}{1}

\section{The Basic Actors}

The abelian and  simple 
real and complex Lie algebras are the  building blocks for all 
real and complex Lie operations\cite{FULHAR,GIL,HEL1} .
The classification of the Lie algebras  with dimensions one, two and three 
- `the basic physical Lie operations (actors)'  - 
gives also the simplest nontrivial examples  for the concepts  
 abelian $\then$ nilpotent $\then$ solvable and simple\cite{LIE13}.  

It makes sense to call the trivial Lie algebra $L=\{0\}$ semisimple, but not
simple. Its group is the trivial group $\exp\{0\}=\{1\}$.

There is one{ complex 1-di\-men\-sio\-nal Lie algebra} $L\cong\C$, it is abelian 
$\p L= [L,L]=\{0\}$. It generates the linear group\footnote{\scriptsize
There is the covariant functor $G\mape \log G$ from Lie groups to their
Lie algebras and, vice versa, $L\mape \exp L$.
The linear group $\GL(\K^n)$ for $\K\in\{\R,\C\}$
has as Lie algebra $\AL(\K^n)$ ($(n\x n)$-matrices).}
 $\GL(\C)$.
 
There are two{ complex 2-di\-men\-sio\-nal Lie algebras}\footnote
{\scriptsize Isomorphies should 
be qualified, e.g. for  a Lie algebra  $L\cong\K^n$ ($\K$-vector space isomorphy).
For  a simpler notation, such  qualifications are omitted -
they should be obvious from the context.}
 $L\cong\C^2$, the abelian
decomposable one and - new for two dimensions -  the  nonabelian, solvable one.
The latter one is a semidirect product\footnote{\scriptsize
denoted as $G_1\sx G_2$ for groups and $L_1\rvec\pl L_2$ for Lie algebras}, isomorphic to the Lie algebra of
 the 1-di\-men\-sio\-nal
affine group.
It is given in the 2nd column  with the bracket in a basis $\{l^1,l^2\}$
and its  faithful adjoint re\-pre\-sen\-ta\-tion

{\scriptsize\begin{eq}{l}
\begin{array}{|c|c|}\hline
\hbox{abelian}&\hbox{solvable}\cr\hline
\log\GL(\C)\cong\C&\log[\GL(\C)\sx\C]\cong\C\rvec\pl\C\cr\hline
[l,l]=0&[l^1,l^2]=l^2\cr
x l\to x\in\C&\psi l^1+xl^2\mape {\scriptsize\pmatrix{0&0\cr
-x&\psi\cr}}\in\AL(\C^2)\cr\hline
\log\D(1)\cong\R&\log[\D(1)\sx\R]\cong\R\rvec\pl\R\cr
\log\U(1)\cong\R&\cr\hline\end{array}
\end{eq}}

\noindent Complex Lie operations have real forms:
$\D(1)=\exp\R$ and $\U(1)=\exp i\R$ are the
connected real 1-di\-men\-sio\-nal Lie groups.
The abelian Lie algebras $x\in\R^n$ formalize physical translations.
In the semidirect real Lie group 
${\scriptsize\pmatrix{1&0\cr -x&e^\psi\cr}}\in \D(1)\sx\R$
the translations are 
acted upon with
 $\D(1)$-dilations $\R\ni x\mape e^\psi x$.

There are three nondecomposable complex 3-dimensional   
Lie algebras $L\cong\C^3$
 - simple, solvable and nilpotent 
(proof in the appendix). They 
have faithful re\-pre\-sen\-ta\-tions for a basis $\{l^1,l^2,l^3\}$
by $(3\x3)$-matrices
$\al_1l^1+\al_2l^2+\al_3l^3\mape\AL(\C^3)$

{\scriptsize\begin{eq}{l}
\begin{array}{|c|c|c|}\hline
\hbox{simple}&\hbox{solvable}&\hbox{nilpotent}\cr\hline
\log\SO(\C^3)&\log[\SO(\C^2)\sx\C^2]&\log\bl H(\C)\cr
&\cong\log\SO(\C^2)\rvec\pl\C^2&\cong\C\rvec\pl\C^2\cr \hline
\begin{array}{rl}
[l^1,l^2]&= l^3\cr
[l^2,l^3]&=l^1 \cr
[l^3,l^1]&=l^2\cr
\end{array}&
\begin{array}{rl}
[l^1,l^2]&= 0\cr
[l^2,l^3]&=l^1 \cr
[l^3,l^1]&=l^2\cr
\end{array}&
\begin{array}{rl}
[l^1,l^2]&= 0\cr
[l^2,l^3]&=0 \cr
[l^3,l^1]&=l^2\cr
\end{array}\cr\hline
 {\scriptsize\left(
\begin{array}{ccc}
0&\al_3&\al_2\cr
-\al_3&0&\al_1\cr
-\al_2&-\al_1&0\cr\end{array}\right)}&
{\scriptsize\left(
\begin{array}{cc|c}
0&\al_3&\al_2\cr
-\al_3&0&\al_1\cr\hline
0&0&0\cr\end{array}\right)}&
{\scriptsize\left(
\begin{array}{cc|c}
0&\al_3&\al_2\cr
0&0&\al_1\cr\hline
0&0&0\cr\end{array}\right)}\cr\hline
\log\SO(3)&\log\SO(2)\rvec\pl\R^2&\log\bl H(1)\cr
\log\SO_0(1,2)&\log\SO_0(1,1)\rvec\pl\R^2&\cong\R\rvec\pl\R^2
\cr\hline
\hbox{rotations}&\hbox{flat Euclidean}&\hbox{Heisenberg}\cr
\hbox{flat Lorentz}&\hbox{flat Poincar\'e}&\cr\hline
\end{array}

\end{eq}}

\noindent Their real forms (2nd last line)
are - up to $\log\SO(3)$ - noncompact. 

All nondecomposable Lie algebras with dimensions 1,2 and 3 have rank 1.
There is one generating linear invariant for the abelian case, 
and one quadratic Casimir invariant
(inverse Killing form) for the simple case 
and its contracted forms 
for the contractions (below). 
The 2-di\-men\-sio\-nal Lie algebra $\R\rvec\pl\R$
has no nontrivial invariant.

The  Lie groups for the 1-di\-men\-sio\-nal real  Lie algebras
act irreducibly  on 1-di\-men\-sio\-nal spaces and, in the selfdual orthogonal
re\-pre\-sen\-ta\-tions, on 2-di\-men\-sio\-nal ones
\begin{eq}{rl}
\hbox{noncompact:}&
\left\{\begin{array}{l}
\D(1)=\exp\R\ni e^\psi,~~
\R\cong\log\D(1)\cong\log \SO_0(1,1)\cr 
\SO_0(1,1)\cong\exp\si_3\R\ni
{\scriptsize\pmatrix{
e^\psi&0\cr
0&e^{-\psi}\cr}}\cong
{\scriptsize\pmatrix{
\cosh\psi&\sinh\psi\cr
\sinh\psi&\cosh\psi\cr}}\in \exp\si_1\R\cr

\end{array}\right.\cr\cr
\hbox{compact:}&
\left\{\begin{array}{l}
\U(1)=\exp i\R\ni e^{i\phi},~~
i\R\cong\log\U(1)\cong\log \SO(2)\cr 
\SO(2)\cong \exp i\si_3\R\ni
{\scriptsize\pmatrix{
e^{i\phi}&0\cr
0&e^{-i\phi}\cr}}\cong
{\scriptsize\pmatrix{
\cos\phi&i\sin\phi\cr
i\sin\phi&\cos\phi\cr}}\in\exp i\si_1\R
\cr
\end{array}\right.
\end{eq}The simply connected totally ordered group
$\D(1)\cong\R$ covers 
$\U(1)\cong\R/\Z$ infinitely often
and is a real form of the complex 1-di\-men\-sio\-nal full Lie group 
$\GL(\C)\cong\SO(\C^2)$ with Lie algebra $D_1$.

The  simple real   Lie structures with dimension 3  add 
 spherical\footnote{\scriptsize
The compact $s$-sphere $\Om^s\cong\SO(1+s)/\SO(s)$
and the noncompact $s$-hyperboloid $\cl Y^s\cong\SO_0(1,s)/\SO(s)$
for $s=1,2,\dots$ pa\-ra\-me\-tri\-ze classes of orthogonal groups.}
and  hyperbolic  degrees 
 of freedom to the abelian $\R$-structures 
\begin{eq}{rll} 
\SO(2)&\cong\Om^1,&\SO_0(1,1)\cong\cl Y^1\cr
\SO(3)&\hbox{rotation group},&\SO(3)/\SO(2)\cong\Om^2\cr
\SO_0(1,2)
&\hbox{flat Lorentz group},&\left\{\begin{array}{rl}
\SO_0(1,2)/\SO(2)&\cong\cl Y^2\cr
\SO_0(1,2)/\SO_0(1,1)&\cong\cl Y^1\x\Om^1\cr\end{array}\right.
\end{eq}The twofold covering groups are (iso)spin
$\SU(2)$ (simply connected) and 
$\SU(1,1)$ ($\Z$-connected) as real forms of the complex 3-di\-men\-sio\-nal 
special Lie group $\SL(\C^2)$ (considered as 3-di\-men\-sio\-nal complex Lie group).

The simple complex  Lie operations, i.e. nonabelian without proper ideal,
have been classified
by Cartan with four main series $\{A,B,C,D\}$  and five exceptional Lie algebras.
Three main series - with the invariance operations for volumes 
$A_r\cong\log\SL(\C^{1+r})$,
for odd dimensional orthogonal structures  $B_r\cong\log\SO(\C^{1+2r})$ and for symplectic 
structures 
$C_r\cong\log\Sp(\C^{2r})$ -  
  start with the same simplest simple
Lie algebra which has{ three} dimensions
\begin{eq}{l}
A_1= B_1= C_1\cong\C^3
\end{eq}The fourth series - for even dimensional orthogonal structures -  starts 
with $D_3\cong\log\SO(\C^6)\cong\log\SL(\C^4)$ 
after  the abelian $D_1\cong\log\SO(\C^2)$ and the semisimple 
$D_2\cong A_1\pl A_1\cong\log\SO(\C^4)$.
All simple Lie algebras are `fused collectives' of several 
$A_1$-isomorphic  building blocks.
The simplest example, used in physics, is the 8-dimensional Lie structure
$\SU(3)$ (flavor or color) where  three 
3-dimensional $A_1$-building blocks, called $I$, $U$ and $V$-spin,
are `fused' in their Cartan subalgebras by the linear dependence 
$I^3+U^3+V^3=0$ to yield the Lie algebra $A_2=\log\SL(\C^3)$.

The simplest real simple structures are descendants of $A_1$,
compact and noncompact (definite and indefinite unitary) 
\begin{eq}{rll}
A^c_1= B^c_1= C^c_1&\cong\log\SU(2)&\cong\log\SO(3)\cong\R^3\cr
A^n_1= B^n_1= C^n_1&\cong\log\SU(1,1)&\cong\log\SO_0(1,2)\cong
\log \SL(\R^2)\cong\R^3
\end{eq}

If each of the  three   compact degrees of freedom in the Lie algebra 
of the spin group $\SU(2)\cong\exp A_1^c$ 
is paired with a noncompact one,
there arises  the  rank 2 simple
 Lie algebra $A_1^c\pl iA_1^c$ as `complexified spin' operations
 for the Lie group 
 $\SL(\C^2)=\exp[A_1^c\pl iA_1^c]$,
   considered as  6-di\-men\-sio\-nal real Lie group
   where three independent rotations are paired with three boosts.
 It is the  twofold cover of the 
 orthochronous Lorentz group  
for 4-di\-men\-sio\-nal Minkowski spacetime
\begin{eq}{l}
A_1^c\pl iA_1^c\cong
\log\SL(\C^2)\cong\log\SO(\C^3)\cong\log\SO_0(1,3)\cong\R^6
\end{eq}

The not simple nondecomposable
real 3-di\-men\-sio\-nal Lie operations are all semidirect groups,
i.e. affine subgroups in $\GL(\R^2)\sx\R^2$
\begin{eq}{rl} 
\SO(2)\sx\R^2&\hbox{Euclidean (flat Galilei)  group}\cr
\SO_0(1,1)\sx\R^2&\hbox{flat Poincar\'e group}\cr
\bl H(1)\cong\R\sx\R^2&\hbox{Heisenberg group}\cr
 \end{eq}They are contraction of $\SO(3)$ and $\SO_0(1,2)$ (below). 
 $\D(1)\sx\R$ is a subgroup of the flat Poincar\'e group. 

The general Euclidean, Lorentz and Poincar\'e groups are
$\SO(s)\sx\R^{s}$, $\SO_0(1,s)$ and $\SO_0(1,s)\sx\R^{1+s}$ respectively, 
for $s=1,2,\dots$.
In the physical names of the operations `flat' is meant as 
`spatially flat', i.e. without
nonabelian space rotations $s=1,2$. 
The Heisenberg group $\bl H(n)=\R^n\sx\R^{1+n}$ with $n$ position-mo\-men\-tum pairs
is looked at  in more detail below.

\section{Heisenberg Lie Algebras and Groups}

The  simplest nonabelian nilpotent Lie operations constitute the
 real 3-di\-men\-sio\-nal 
{Heisenberg Lie algebra} $\log\bl H(1)=\bl h(1)$ for one
position-mo\-men\-tum pair $(\bl x,\bl p)$ and its 
bracket $\bl I$, a basis for the centrum
\begin{eq}{rl}
\bl h(1)&=\{q\bl x+y\bl p+t\bl I\mid q,y,t\in\R\}\cong\R^3\cr
\hbox{with}&[\bl x,\bl p]=\bl I,~[\bl I,\bl x]=0=[\bl I,\bl p]\cr
\end{eq}The central action operator $\bl I$ is no number, e.g. not
the imaginary unit $i$. 
$\bl h(1)$ is a semidirect product 
\begin{eq}{l}
\bl h(1)=\R\bl x\rvec\pl[\R\bl p+\R\bl I] 

:~~[\R\bl x,\R\bl p+\R\bl I]\sub\R\bl I 
\end{eq}The position acts upon the ideal spanned by
$\{\bl p,\bl I\}$,
not by $\{\bl p,\bl x\}$. 
Here, and everywhere, the roles of the position 
$\bl x$ and the
mo\-men\-tum $\bl p$  operations  can be exchanged.

$\bl h(1)$  has a 3-di\-men\-sio\-nal
faithful  re\-pre\-sen\-ta\-tion by nilpotent
matrices 
\begin{eq}{rl}
\bl h(1)\ni q\bl x+y\bl p+t\bl I
&\mape 
{\scriptsize\pmatrix{
0&q&t\cr
0&0&y\cr
0&0&0\cr}}\in \AL(\R^3)\cr
\end{eq}By exponentiation there arises the{ Heisenberg group} $\bl H(1)$ 
with the Weyl product\footnote{\scriptsize
In analogy to the commutator ideal of a Lie algebra,
there is the normal commutator subgroup
$(G,G)=\{gk (kg)^{-1}\mid g,k\in G\}$.} expressing the noncommutativity of
position and mo\-men\-tum operators
\begin{eq}{rl}
\bl H(1)\ni 
e^{ q\bl x+ y \bl p+ t\bl I}&\mape
{\scriptsize\pmatrix{
1&q&t+qy\cr
 0&1&y\cr
 0 & 0&1\cr}}\in\SL(\R^3)\cr
\hbox{ Weyl product:}&e^{ q \bl x}e^{ y \bl p}={\scriptsize\pmatrix{
1&q&qy\cr
 0&1&y\cr
 0& 0&1\cr}}=e^{ q y \bl I}e^{y \bl p}e^{ q\bl x}\cr

\hbox{commutator group:}&
(\bl H(1),\bl H(1))=e^{\R \bl I}:
~e^{ q\bl x}e^{ y\bl p}e^{-q\bl x}e^{-q\bl p} =e^{ qy \bl I}\cr
\end{eq}

$\bl H(1)$ leads to
the chains with abelian normal subgroups
\begin{eq}{l}
\bl H(1)=\{e^{ q\bl x+ y\bl p+ t\bl I}\}\supnoteq 
\begin{array}{c}
\{e^{ y\bl p+ t \bl I}\}\cong\R^2\cr
\{e^{  q\bl x+ t\bl  I}\}\cong\R^2\cr\end{array}
\supnoteq \centr\bl H(1)=\{e^{ t \bl I}\}\cong\R
\end{eq}Correspondingly, $\bl H(1)$ can be considered either as
 semidirect extension\cite{LIE13} or as central extension  with the exact
sequences\footnote{\scriptsize
An extended  vector space  or group  or Lie algebra $G$ is defined by 
the injection-projection structure 
 $N\stackrel\io\map G\stackrel\pi\map H$
with image $\io=$ kernel $\pi$, i.e. $H\cong G/N$.} 
\begin{eq}{rcl}
\R^2\stackrel{\io_s}\map&\bl H(1)\cong\R\sx\R^2&\stackrel{\pi_s}\map\R\cr
\R\stackrel{\io_c}\map&\bl H(1)\cong\R\lvec\odot\R^2&\stackrel{\pi_c}\map\R^2\cr
\end{eq}In the semidirect group product  
\begin{eq}{l}
 \bl H(1)\cong
\R^2\o\R=\R\sx \R^2 \hbox{ with }\left\{\begin{array}{rl}
\R&\cong\{e^{ q\bl x}\mid q\in\R\}\cr
\R^2&\cong\{e^{ y \bl p+ t \bl I}\mid y, t\in\R\}\cr
\end{array}\right.\cr
\end{eq}illustrated   in the $(3\x3)$-matrix re\-pre\-sen\-ta\-tions
\begin{eq}{l}
{\scriptsize\left(\begin{array}{cc|c}
1&q&t+qy\cr
 0&1&y\cr\hline
 0&0&1\cr\end{array}\right)}
={\scriptsize\left(\begin{array}{cc|c}
1&0&t+qy\cr
0&1&y\cr\hline
 0&0&1\cr\end{array}\right)}
{\scriptsize\left(\begin{array}{cc|c}
1&q&0\cr
0&1&0\cr\hline
0& 0&1\cr\end{array}\right)}
\end{eq}the homogeneous group
with the position $\bl x$ 
acts on  the abelian normal subgroup $\R^2$ by inner 
automorphisms  
\begin{eq}{l}
\R\sx\R^2\map\R^2:~~
\left\{\begin{array}{rl}
e^{q\bl x}\o e^{y\bl p+t\bl I}\o e^{-q\bl x}&= e^{y\bl p+(qy+t)\bl I}\cr
{\scriptsize\pmatrix{1&q\cr 0&1\cr}}{\scriptsize\pmatrix{t\cr y\cr}}&=
{\scriptsize\pmatrix{ t+qy\cr y\cr}}\end{array}\right.\cr
\end{eq}

The  adjoint re\-pre\-sen\-ta\-tion\footnote{\scriptsize
The adjoint re\-pre\-sen\-ta\-tion of a Lie algebra acts on
itself  $\ad:L\map\AL(L)$
with $\ad m(l)=[m,l]$. The adjoint re\-pre\-sen\-ta\-tion of a 
 Lie group on its Lie algebra
$\Ad:G\map\GL(L)$ goes via $\Ad g(l)\sim glg^{-1}$. 
The adjoint action of a group on itself is by inner automorphisms
$\Int g:G\map G$, $\Int g(k)=gkg^{-1}$.}
has commuting position and mo\-men\-tum 
$[\ad \bl x,\ad\bl p]=0$ - the image
is the classical position-mo\-men\-tum Lie algebra   
\begin{eq}{ll}
\ad:\bl h(1)\map\AL(\R^3),&\Ad:\bl H(1)\map\SL(\R^3)\cr
\ad(q\bl x+y\bl p+ t\bl I)=
{\scriptsize\left(\begin{array}{cc|c}
0&0&-y\cr
0&0&q\cr\hline
0&0&0\cr
\end{array}\right)},&
\Ad e^{q \bl x+y \bl p+ t\bl I}=
{\scriptsize\left(\begin{array}{cc|c}
1&0&-y\cr
0&1&q\cr\hline
0&0&1\cr\end{array}\right)}\cr
\ad\bl h(1)\cong\{\R \bl x+\R \bl p\},~\ad \bl I=0,&\Ad \bl I=\bl 1_3\cr

\end{eq}$\bl 1_n=\id_V$ denotes the identity operation on a vector space $V\cong\K^n$.

Also the general Heisenberg Lie algebra with $n$ position mo\-men\-tum pairs, 
$a=1,\dots,n$ 
\begin{eq}{l}
\bl h(n)\cong\R^{1+2n}:~~
[\bl x_a,\bl p^b]=\de_a^b\bl I,~~
[\bl x_a,\bl I]=0
=[\bl I,\bl p^b]
\end{eq}is nilcubic with its centrum as commutator ideal
\begin{eq}{l}
[\bl h(n),\bl h(n)]=\R \bl I=\centr\bl h(n),~~
[[\bl h(n),\bl h(n)],\bl h(n)]=\{0\}
\end{eq}It has a faithful
re\-pre\-sen\-ta\-tion by  $((2+n)\x(2+ n))$-matrices
\begin{eq}{l}
\bl h(n)\ni q^a\bl x_a+ y_b\bl p^b+t\bl I\mape
{\scriptsize\left(\begin{array}{c|c|c}
0&q^a&t\cr\hline
0&0&y_b\cr\hline
0&0&0\cr\end{array}\right)}\in{\scriptsize\left(\begin{array}{c|c|c}
0&\R^n&\R\cr\hline
0&0&\R^n\cr\hline
0&0&0\cr\end{array}\right)}\subnoteq
\AL(\R^{2+n})
\end{eq}It is a semidirect
product of the $n$-dimensional  abelian positions acting upon
the $n$-dimensional mo\-men\-ta and the central operator
\begin{eq}{l}
\bl h(n)=\R^n\rvec \pl[\R^n\pl\R]:~~
[\R\bl x_a,\R\bl p^b+\R\bl I]\sub\R\bl I 
\end{eq}

The{ affine  Heisenberg Lie algebra} ${\bf ah}(n)$  includes, in addition, 
all linear transformations $f\in\AL(\R^n)$
of the position-mo\-men\-tum pairs -
in the matrix re\-pre\-sen\-ta\-tion
\begin{eq}{l}
\bl{ah}(n)\ni
{\scriptsize\left(\begin{array}{c|c|c}
0&q^a&t\cr\hline
0&f_b^a&y_b\cr\hline
0&0&0\cr\end{array}\right)}\in
{\scriptsize\left(\begin{array}{c|c|c}
0&\R^n&\R\cr\hline
0&\AL(\R^n)&\R^n\cr\hline
0&0&0\cr\end{array}\right)}\subnoteq \AL(\R^{2+n})
\end{eq}with the Lie-bracket involving 
the dual product $\dprod qy$ 
of  position with mo\-men\-tum and 
the linear transformations
acting on position and mo\-men\-tum
\begin{eq}{l}
[{\scriptsize\left(\begin{array}{c|c|c}
0&q_1&t_1\cr\hline
0&f_1&y_1\cr\hline
0&0&0\cr\end{array}\right)},
{\scriptsize\left(\begin{array}{c|c|c}
0&q_2&t_2\cr\hline
0&f_2&y_2\cr\hline
0&0&0\cr\end{array}\right)}]
=
{\scriptsize\left(\begin{array}{c|c|c}
0&f^T_2(q_1)-f^T_1(q_2)&\dprod{q_1}{y_2}-\dprod{q_2}{y_1}\cr\hline
0&[f_1,f_2]&f_1(y_2)-f_2(y_1)\cr\hline
0&0&0\cr\end{array}\right)}\cr
\cr
\hbox{with }\dprod qy=q^ay_a,~~
f(y)_b=f^a_by_a,~~f^T(q)^a=f^a_bq^b

\end{eq}The  Heisenberg Lie algebra is the nilradical of
the  affine Heisenberg Lie algebra
\begin{eq}{rl}
[\bl{ah}(n), \bl h(n)]&\sub \bl h(n)\cr

[{\scriptsize\left(\begin{array}{c|c|c}
0&\R^n&\R\cr\hline
0&\AL(\R^n)&\R^n\cr\hline
0&0&0\cr\end{array}\right)},
{\scriptsize\left(\begin{array}{c|c|c}
0&\R^n&\R\cr\hline
0&0&\R^n\cr\hline
0&0&0\cr\end{array}\right)}]
&\sub
{\scriptsize\left(\begin{array}{c|c|c}
0&\R^n&\R\cr\hline
0&0&\R^n\cr\hline
0&0&0\cr\end{array}\right)}

\end{eq}Therewith the extended Heisenberg Lie algebra is a double semidirect
product 
\begin{eq}{l}
\bl{ah}(n)=\AL(\R^n)\rvec\pl \bl h(n)=
 \AL(\R^n)\rvec\pl [\R^n\rvec\pl[\R^n\pl\R]]
\end{eq}

\section{Contractions}

In general, contractions of simple
Lie operations go with a `flattening' of degrees of freedom.
In physics, this is related to a trivialization of units.
The prototype Wigner-In\"on\"u contraction from the   
Lorentz to the Galilei group\cite{INWIG}
`flattens' the boosts by  trivializing the speed of light
${1\over c}\to 0$. In the opposite procedure,
a contracted group is expanded by `flexing' flat degrees of freedom. 
An expansion resuscitates a mute unit.
Contractions work with units, operationally formalized with dilations.

The not simple nondecomposable
real 3-di\-men\-sio\-nal Lie operations 
are all contractions of the simplest simple Lie operations.
Their Lie algebras are given with a basis,
 a { defining re\-pre\-sen\-ta\-tion} 
in the endomorphism algebra $\AL(\C^3)$ (complex $(3\x3)$-matrices)  - 
for the simple groups $\SO(3)$ and $\SO_0(1,2)$ the adjoint
re\-pre\-sen\-ta\-tions - and the invariant Casimir element $\bl C$
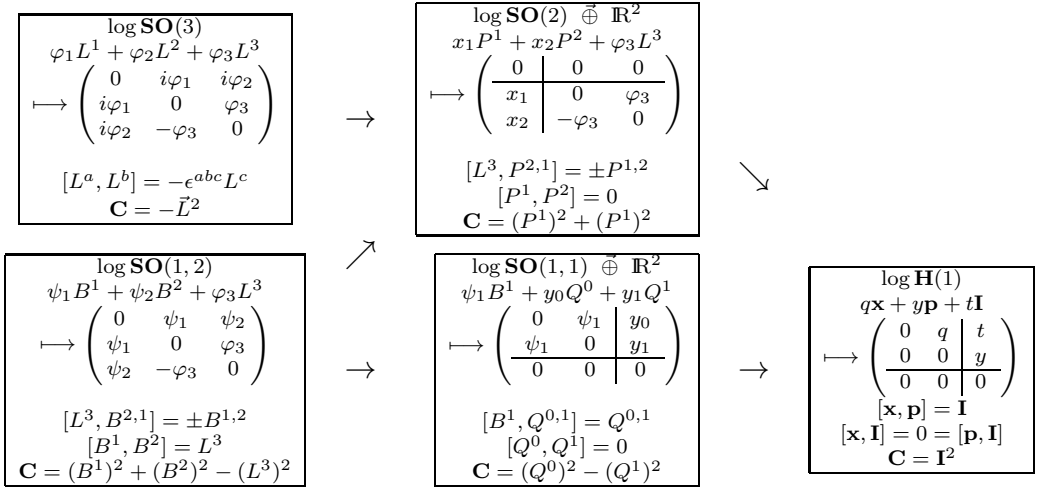
\begin{eq}{ccrcr}

{\scriptsize\begin{array}{|c|}\hline
 \log \SO(3)\cr
\phi_1L^1+\phi_2L^2
+\phi_3L^3\cr
\mape{\scriptsize\pmatrix{
0&i\phi_1&i\phi_2\cr
i\phi_1&0&\phi_3\cr
i\phi_2&-\phi_3&0\cr}}\cr\cr
[L^a,L^b]=-\ep^{abc}L^c\cr
\bl C=-\rvec L^2\cr\hline
\end{array}}
&\to&
{\scriptsize\begin{array}{|c|}\hline
\log \SO(2)\rvec\pl\R^2\cr
x_1P^1+x_2P^2
+\phi_3L^3\cr
\mape
{\scriptsize\left(\begin{array}{c|cc}
0&0&0\cr\hline
x_1&0&\phi_3\cr
x_2&-\phi_3&0\cr\end{array}\right)}
\cr\cr
[L^3,P^{2,1}]=\pm P^{1,2}\cr
[P^1,P^2]=0\cr
\bl C=(P^1)^2+(P^1)^2\cr
\hline
\end{array}}&\begin{array}{c}
\cr\cr\cr\searrow
\end{array}&\cr

{\scriptsize\begin{array}{|c|}\hline
\log \SO(1,2)\cr
\psi_1B^1+\psi_2B^2
+\phi_3L^3\cr
\mape
{\scriptsize\pmatrix{
0&\psi_1&\psi_2\cr
\psi_1&0&\phi_3\cr
\psi_2&-\phi_3&0\cr}}\cr\cr
[L^3,B^{2,1}]=\pm B^{1,2}\cr
[B^1,B^2]=L^3\cr
\bl C=(B^1)^2+(B^2)^2-(L^3)^2\cr
\hline
\end{array}}&\begin{array}{c}
\nearrow\cr\cr\cr
\to\cr\cr\cr\cr\end{array}
&

{\scriptsize\begin{array}{|c|}\hline
\log \SO(1,1)\rvec\pl\R^2\cr
\psi_1B^1
+y_0Q^0+y_1Q^1
\cr
\mape
{\scriptsize\left(\begin{array}{cc|c}
0&\psi_1&y_0\cr
\psi_1&0&y_1\cr\hline
 0& 0&0\cr\end{array}\right)}\cr\cr
[B^1,Q^{0,1}]=Q^{0,1}\cr
[Q^0,Q^1]=0\cr
\bl C=(Q^0)^2-(Q^1)^2\cr\hline
\end{array}}

&\to &
{\scriptsize\begin{array}{|c|}\hline
\log \bl H(1)\cr
 q\bl x+y\bl p+t\bl I\cr
\mape {\scriptsize\left(\begin{array}{cc|c}
0&q&t\cr
 0&0&y\cr\hline
 0&0&0\cr
 \end{array}\right)}\cr
[\bl x,\bl p]=\bl I\cr
[\bl x,\bl I]=0=[\bl p,\bl I]\cr
\bl C=\bl I^2\cr\hline
\end{array}}
\cr

\cr

\end{eq}From the two simple groups (left), 
3-rotations and $(1,2)$-Lorentz group, related to each other by the 
spherical-hyperbolic, i.e.
compact-noncompact, i.e. imaginary-real  exchange of two operations
$i\phi_{1,2}\lrmap \psi_{1,2}$,
 there leads  one contraction
to the 2-Euclidean and (or) to the $(1,1)$-Poincar\'e group. From 
these contracted groups there leads  a 2nd contraction
to the double contracted Heisenberg group which can also 
be reached  directly by a
central contraction.

\subsection{Contractive Units and  Dilations}

The simple Lie algebra pa\-ra\-me\-tri\-zations above
with  faithful adjoint $(3\x3)$-matrix re\-pre\-sen\-ta\-tions
are for a diagonal invariant  metric
$\bl1_3$ and $\eta$, i.e. for the Killing forms in an  orthonormal basis.
Dilation transformations from the triad manifold
$d\in\D(1)^3\in \GL(\R^3)/\SO(3)$ and $ d\in\D(1)^3\in \GL(\R^3)/\SO_0(1,2)$
introduce three units 
$\al_{1,2,3}>0$ for the three operations. The units can be visualized as 
lengths for the three axes of a  metrical 2-ellipsoid $d\bl1_3 d^T$ 
for $\SO(3)$ and 
of a metrical 2-hyperboloid $d\eta d^T$ for $\SO_0(1,2)$
\begin{eq}{rc}
\SO(3):~d\bl1_3 d^T=&{\scriptsize\pmatrix{
\al_1&0&0\cr
0&\al_2&0\cr
0&0&\al_3&\cr}}
{\scriptsize\pmatrix{
1&0&0\cr0&1&0\cr0&0&1\cr}}
{\scriptsize\pmatrix{
\al_1&0&0\cr
0&\al_2&0\cr
0&0&\al_3&\cr}}\cr
&={\scriptsize\pmatrix{
\al^2_1&0&0\cr
0&\al^2_2&0\cr
0&0&\al^3_3&\cr}}\hfill\cr\cr
\SO_0(1,2):~d\eta d^T=&{\scriptsize\pmatrix{
\al_1&0&0\cr
0&\al_2&0\cr
0&0&\al_3&\cr}}
{\scriptsize\pmatrix{
1&0&0\cr0&-1&0\cr0&0&-1\cr}}
{\scriptsize\pmatrix{
\al_1&0&0\cr
0&\al_2&0\cr
0&0&\al_3&\cr}}\cr
&={\scriptsize\pmatrix{
\al^2_1&0&0\cr
0&-\al^2_2&0\cr
0&0&-\al^3_3&\cr}}\hfill
\end{eq}The related inner automorphisms  $L\ni l\mape dld^{-1}$ 
give
renormalized Lie algebra re\-pre\-sen\-ta\-tions
which leave invariant the dilation transformed metric
 - for $\SO_0(1,2)$ 
\begin{eq}{rl}
dld^{-1}&={\scriptsize\pmatrix{
\al_1&0&0\cr
0&\al_2&0\cr
0&0&\al_3&\cr}}
{\scriptsize\left(\begin{array}{ccc}
0&\psi_1&\psi_2\cr
\psi_1&0&\phi_3\cr
\psi_2&-\phi_3&0\cr
\end{array}\right)}
{\scriptsize\pmatrix{
{1\over \al_1}&0&0\cr
0&{1\over \al_2}&0\cr
0&0&{1\over \al_3}&\cr}}\cr
&=
{\scriptsize\left(\begin{array}{ccc}
0&{\al_1\over\al_2}\psi_1&{\al_1\over\al_3}\psi_2\cr
{\al_2\over\al_1}\psi_1&0&{\al_2\over\al_3}\phi_3\cr
{\al_3\over\al_1}\psi_2&-{\al_3\over\al_2}\phi_3&0\cr\end{array}\right)}

\end{eq}and - with
$\psi_{1,2}\mape i\phi_{1,2}$  - for $\SO(3)$.

\subsection{Simple Contractions}

By renormalization of the Lie algebra basis 
and a renaming of the parameters,
e.g. ${\al_1\over\al_3}\psi_2=y_0$,  one obtains the
contraction of the flat Lorentz group $\SO_0(1,2)$ to the flat Poincar\'e group
\begin{eq}{rl}
{\scriptsize\left(\begin{array}{cc|c}
0&{\al_1\over\al_2}\psi_1&{\al_1\over\al_3}\psi_2\cr
{\al_2\over\al_1}\psi_1&0&{\al_2\over\al_3}\phi_3\cr\hline
{\al_3\over\al_1}\psi_2&-{\al_3\over\al_2}\phi_3&0\cr\end{array}\right)}
={\scriptsize\left(\begin{array}{cc|c}
0&{\al_1\over\al_2}\psi_1& y_0\cr
{\al_2\over\al_1}\psi_1&0&y_1\cr\hline
 {\al^2_3\over\al^2_1}y_0&-{\al^2_3\over\al^2_2}y_1& 0\cr\end{array}\right)}
&\to{\scriptsize\left(\begin{array}{cc|c}
0&\psi_1&y_0\cr
\psi_1&0&y_1\cr\hline
0&0&0\cr\end{array}\right)}
\cr
\cr
 \al^2_3\to 0, ~\al_1=\al_2 \hbox{ with finite }y_{0,1}&
\end{eq}Therewith the  Casimir invariant (inverse Killing form) is contracted to
\begin{eq}{l}
\bl C=(B^1)^2+(B^2)^2-(L^3)^2\to
(Q^0)^2-(Q^1)^2
\end{eq}i.e. the metrical 2-hyperboloid degenerates  to a metrical
1-hyperbola. Anticipation of the contraction
 leads to the pa\-ra\-me\-tri\-zation of metrical hyperboloid and
invariance group involving 
one{ contractive unit} $\ell$ as one main axis
\begin{eq}{rl}
{\scriptsize\left(\begin{array}{cc|c}
\al^2_1&0&0\cr
0&-\al^2_2&0\cr\hline
0&0&-\al^3_3\cr
\end{array}\right)}
&=
{\scriptsize\left(\begin{array}{cc|c}
1&0&0\cr
0&-1&0\cr\hline
0&0&- \ell^2\cr
\end{array}\right)}
\cr
\log\SO_0(1,2)\ni {\scriptsize\left(\begin{array}{cc|c}
0&{\al_1\over\al_2}\psi_1& y_0\cr
{\al_2\over\al_1}\psi_1&0&y_1\cr\hline
 {\al^2_3\over\al^2_1}y_0&-{\al^2_3\over\al^2_2}y_1& 0\cr\end{array}\right)}

&=

{\scriptsize\left(\begin{array}{cc|c}
0&\psi_1& y_0\cr
\psi_1&0& y_1\cr\hline
  \ell^2y_0&-\ell^2 y_1&0\cr
\end{array}\right)}\cr
\ell^2\to 0:~~\SO_0(1,2)\to
&\SO_0(1,1)\sx\R^2\cr
\hbox{with }{\scriptsize\pmatrix{
\cosh\psi_1&\sinh\psi_1\cr
\sinh\psi_1&\cosh\psi_1\cr}}&\in\SO_0(1,1)\subnoteq\GL(\R^2)\cr
\end{eq}

The Wigner-In\"on\"u contraction from the flat Lorentz group to the Euclidean 
(flat Galilei) group is analogous
\begin{eq}{rl}
{\scriptsize\left(\begin{array}{c|cc}
0&{\al_1\over\al_2}\psi_1&{\al_1\over\al_3}\psi_2\cr\hline
{\al_2\over\al_1}\psi_1&0&{\al_2\over\al_3}\phi_3\cr
{\al_3\over\al_1}\psi_2&-{\al_3\over\al_2}\phi_3&0\cr\end{array}\right)}
={\scriptsize\left(\begin{array}{c|cc}
0&{\al^2_1\over\al^2_2}x_1&{\al^2_1\over\al^3_3}x_2\cr\hline
x_1&0&{\al_2\over\al_3}\phi_3\cr
x_2&-{\al_3\over\al_2}\phi_3&0\cr\end{array}\right)}
&\to{\scriptsize\left(\begin{array}{c|cc}
0&0&0\cr\hline
x_1&0&\phi_3\cr
x_2&-\phi_3&0\cr\end{array}\right)}\cr\cr
 \al^2_1\to 0,~\al_2=\al_3 \hbox{ with finite  }&x_{1,2}\cr
 \bl C=(B^1)^2+(B^2)^2-(L^3)^2&\to
(P^1)^2+(P^2)^2

\end{eq}The metrical 2-hyperboloid degenerates to a metrical
circle (1-sphere). The  contractive unit $c$ is the speed of light
\begin{eq}{rl}

{\scriptsize\left(\begin{array}{c|cc}
\al^2_1&0&0\cr\hline
0&-\al^2_2&0\cr
0&0&-\al^3_3\cr\end{array}\right)}

&={\scriptsize\left(\begin{array}{c|cc}
{1\over c^2}&0&0\cr\hline
0&-1&0\cr
0&0&-1\cr\end{array}\right)}
\cr
\log\SO_0(1,2)\ni {\scriptsize\left(\begin{array}{c|cc}
0&{\al^2_1\over\al^2_2}x_1&{\al^2_1\over\al^3_3}x_2\cr\hline
x_1&0&{\al_2\over\al_3}\phi_3\cr
x_2&-{\al_3\over\al_2}\phi_3&0\cr\end{array}\right)}
&=
{\scriptsize\left(\begin{array}{c|cc}
0&{1\over c^2}x_1&{1\over c^2}x_2\cr\hline
x_1&0&\phi_3\cr
x_2&-\phi_3&0\cr\end{array}\right)}\cr
{1\over c^2}\to 0:~~
\SO_0(1,2)\to &\SO(2)\sx\R^2\cr
\hbox{with }{\scriptsize\pmatrix{
\cos\phi_3&\sin\phi_3\cr
-\sin\phi_3&\cos\phi_3\cr}}&\in\SO(2)\subnoteq\GL(\R^2)\cr

\end{eq}

\subsection{Double Contraction}

The transition from   Poincar\'e or Euclid
  to Heisenberg involves a 2nd contraction
 from `hyperbolic' or spherical to flat
\begin{eq}{lcrrll}
&\nearrow &\SO_0(1,1)\sx \R^2&\hbox{then}&\SO_0(1,1)&\to\R\cr
\SO_0(1,2)&&&&&\cr
&\searrow&&&&\cr
& &\SO(2)\sx \R^2&\hbox{then}&\SO(2)&\to\R\cr
\SO(3)&\nearrow&&&&
\end{eq}E.g., the double contraction from simple $\SO_0(1,2)$ to Heisenberg $\bl H(1)$
(Segal contraction\cite{SEG})
\begin{eq}{rl}
{\scriptsize\left(\begin{array}{cc|c}
0&{\al_1\over\al_2}\psi_1&{\al_1\over\al_3}\psi_2\cr
{\al_2\over\al_1}\psi_1&0&{\al_2\over\al_3}\phi_3\cr\hline
{\al_3\over\al_1}\psi_2&-{\al_3\over\al_2}\phi_3&0\cr\end{array}\right)}
=
{\scriptsize\left(\begin{array}{cc|c}
0& q& t\cr
{\al_2^2\over\al_1^2} q&0&y\cr\hline
{\al_3^2\over\al_1^2} t&- {\al_3^2\over\al_2^2}y&0\cr
\end{array}\right)}
&\to{\scriptsize\left(\begin{array}{cc|c}
0&q&t\cr
 0&0&y\cr\hline
 0&0&0\cr\end{array}\right)}
\cr
\cr
\al_3\to 0,~~{1\over \al_1}\to 0,~ \al_2=1 \hbox{ with finite }q,y,t&\cr
 
\bl C=(B^1)^2+(B^2)^2-(L^3)^2&\to \bl I^2

\end{eq}can be performed  with  two contractive units 
$\ell^2,{1\over c^2}$
\begin{eq}{rl}
{\scriptsize\left(\begin{array}{cc|c}
\al^2_1&0&0\cr
0&-\al^2_2&0\cr\hline
0&0&-\al^3_3\cr\end{array}\right)}

&=
{\scriptsize\left(\begin{array}{cc|c}
c^2&0&0\cr
0&-1&0\cr\hline
0&0&-\ell^2\cr\end{array}\right)}

\cr
\log\SO_0(1,2)\ni{\scriptsize\left(\begin{array}{cc|c}
0& q& t\cr
{\al_2^2\over\al_1^2} q&0&y\cr\hline
{\al_3^2\over\al_1^2} t&- {\al_3^2\over\al_2^2}y&0\cr
\end{array}\right)}

&={\scriptsize\left(\begin{array}{cc|c}
0&q& t\cr
{1\over c^2}q&0& y\cr\hline
{\ell^2\over c^2} t& -\ell^2y &0\cr\end{array}\right)}\cr
 \ell^2,{1\over c^2}\to0:~~
\SO_0(1,2)\to& \bl H(1)\cong\R\sx\R^2\cr
\hbox{with }{\scriptsize\pmatrix{
1&q\cr
0&1\cr}}&\in\R\subnoteq\GL(\R^2)\cr

\end{eq}A 1-step contraction
uses  one  contractive unit 
$\mu^2\to 0$
\begin{eq}{rl}
{\scriptsize\left(\begin{array}{cc|c}
\al^2_1&0&0\cr
0&-\al^2_2&0\cr\hline
0&0&-\al^3_3\cr\end{array}\right)}

&={\scriptsize\left(\begin{array}{cc|c}
1&0&0\cr
0&-\mu^2&0\cr\hline
0&0&-\mu^4\cr\end{array}\right)}

\cr
\log\SO_0(1,2)\ni{\scriptsize\left(\begin{array}{cc|c}
0& q& t\cr
{\al_2^2\over\al_1^2} q&0&y\cr\hline
{\al_3^2\over\al_1^2} t&- {\al_3^2\over\al_2^2}y&0\cr
\end{array}\right)}

&={\scriptsize\left(\begin{array}{cc|c}
0&q& t\cr
\mu^2q&0& y\cr\hline
\mu^4 t& -\mu^2 y &0\cr\end{array}\right)}\cr

\end{eq}

\subsection{Semidirect and Central Contractions}
 
Any subgroup $H$ in a finite dimensional Lie  group $G$ allows 
vector space decompositions of the Lie algebra
\begin{eq}{l}
\log G=\log H\pl K,~~K\cong\log G/\log H
\hbox{ with }
\left\{\begin{array}{rl}
[\bl h_1,\bl h_2]&=\bl h_3\cr
[\bl h_1,\bl k_1]&=\bl h_2+\bl k_2\cr
[\bl k_1,\bl k_2]&= \bl h+\bl k_3\cr\end{array}\right.
\end{eq}A dilation transformation of the complementary vector subspaces
\begin{eq}{l}
\begin{array}{l}
\mu\nu\ne0,~\bl h={1\over\nu}h,~~\bl k={1\over\mu}k\cr
\log G=\log H(\nu)\pl K(\mu)
\end{array}~~~~
\left\{\begin{array}{rl}
[h_1,h_2]&=\nu h_3\cr 
[h_1,k_1]&=\mu h_2+\nu k_2\cr
[k_1,k_2]&= {\mu^2\over \nu} h+\mu k_3\end{array}\right.
\end{eq}can be used for
the{ semidirect contraction} with the
Lie subalgebra $\log H$ acting on a vector space $K$
\begin{eq}{rl}
\nu=1,~\mu\to 0:&\log G\to \log H\rvec\pl K,~~
\left\{\begin{array}{rl}
[\log H,\log H]&\sub \log H\cr 
[\log H,K]&\sub K\cr
[K,K]&=\{0\}\end{array}\right.\cr
\hbox{e.g.}&\hskip6mm \log\SO(3)\to\log \SO(2)\rvec\pl\R^2\cr 
&\log\SO_0(1,2)\to\log \SO_0(1,1)\rvec\pl\R^2\cr 
\end{eq}With  related dilations 
the{ central contraction} gives a central Lie subalgebra $\log H$.
The complementary space $K$ has a Lie bracket in $\log H$  
\begin{eq}{rl}
\nu=\mu^2:&
\left\{\begin{array}{rl}
[h_1,h_2]&=\mu^2 h_3\cr 
[h_1,k_1]&=\mu h_2+\mu^2 k_2\cr
[k_1,k_2]&= h+\mu k_3\end{array}\right.\cr
\mu\to 0:&
\begin{array}{rl}
\log G\to &\log H~~\lvec\odot~~ K\cr
\log H\sub&\centr[\log H~~\lvec\odot~~ K]\end{array}
,~~\left\{\begin{array}{rl}
[\log H,\log H]&=\{0\}\cr 
[\log H,K]&=\{0\}\cr
[K,K]&\sub \log H\end{array}\right.\cr
\hbox{e.g.}&\log\SO(3),\log\SO_0(1,2)\to\R~~\lvec\odot~~\R^2\cr 

\end{eq}In the Heisenberg example $\R^2\subnoteq\log\bl H(1)$ is spanned by 
$\{\bl p,\bl x\}$ for the central contraction 
in contrast to  $\{\bl x,\bl I\}$ for the semidirect contraction.

The diagonalizable quadratic  Casimir element for a semisimple 
Lie algebra (Killing metric of the adjoint re\-pre\-sen\-ta\-tion) is contracted 
to a bilinear vector space form in the semidirect
contraction and to a central subalgebra  form  in the central one
\begin{eq}{rl}
\log G:&\bl C=
 \rvec{\bl  h}\ox \rvec {\bl  h}+ \rvec{\bl k}\ox \rvec{\bl k} 
={ \rvec h\ox \rvec h\over\nu^2}+{ \rvec k\ox \rvec k\over\mu^2}\cr
\log G\to \log H\rvec\pl K:&
\mu^2\bl C\to   \rvec k\ox  \rvec k\cr
\log G\to \log H~\lvec\odot~ K:&
\mu^4\bl C\to   \rvec h\ox \rvec h\cr
\end{eq}

\section{Hilbert Representations}

Groups carry `in themselves' the structure of `their' re\-pre\-sen\-ta\-tion spaces:
Any set $S$ where a group $G$ acts on, is a disjoint union of
$G$-orbits $G\m v$, $v\in S$, which are irreducible $G$-sets.
An orbit is 
 isomorphic to a subgroup class $G\m v\cong  G/H$, i.e. to
 the $G$-operations  up to 
 the fixgroup (Wigner's 'little' groups) $H=G_v$
for the $G$-action. Therefore:
The coset spaces $\{G/H\mid H\sub G\}$ constitute - up to isomorphy -
 the irreducible sets with $G$-action.
 
 For  group re\-pre\-sen\-ta\-tions on vector spaces, linearity has to be taken into
 account (more below). All  
real Lie groups (locally compact) define `their'  Hilbert spaces with 
complex re\-pre\-sen\-ta\-tions,   compact Lie groups
have only Hilbert re\-pre\-sen\-ta\-tions. In physics, with Born, the  
scalar product of the Hilbert
spaces acted upon is interpreted in terms of  `probability amplitudes'.
Therewith, physical Lie operations carry  their 
probability interpretation in their own structure.

\subsection{Some General Remarks}

First some facts about  re\-pre\-sen\-ta\-tions
which, in a more detailed and exact formulation, can be
found in the literature\cite{LIE13,FOL,KIR}:
Group (Lie algebra) re\-pre\-sen\-ta\-tions have a normal subgroup
(an ideal) as kernel, i.e. a group (Lie algebra)
without normal subgroup (ideal) has  faithful, i.e. injective, or 
trivial  re\-pre\-sen\-ta\-tions. For the abelian and simple groups 
with  the `basic physical Lie operations' 
the nontrivial unfaithful re\-pre\-sen\-ta\-tions
are characterized by  the discrete normal 
subgroups $N$ - they are faithful for the
 quotient groups $G/N$
 
{\scriptsize \begin{eq}{l}
\begin{array}{|c|c||c|c|}\hline
G&\dim_\R G&N&G/N\cr\hline\hline
\D(1)\cong\R&1&e^\Z\cong \Z&\U(1)\cr\hline
\SU(2)&3&\{\pm \bl1_2\}&\SO(3)\cr\hline
\SU(1,1)&3&\{\pm \bl1_2\}&\SO_0(1,2)\cr\hline
\end{array}\cr
\end{eq}}

\noindent For the affine group in one dimension and  for the
 3-dimensional  contracted  groups 
the translations are continuous normal subgroups 

{\scriptsize\begin{eq}{l}
\begin{array}{|c|c||c|c|}\hline
G&\dim_\R G&N&G/N\cr\hline\hline
\D(1)\sx\R&2&
\begin{array}{c}
\R\cr 
\end{array}
&\begin{array}{c}
\D(1)\cr
\end{array}\cr
\hline
\SO(2)\sx\R^2&3&\R^2&\SO(2)\cr\hline

\SO_0(1,1)\sx\R^2&3&
\begin{array}{c}
\R^2\cr
\end{array}
&\begin{array}{c}
\SO_0(1,1)\cr
\end{array}\cr
\hline

\bl H(1)=\R\sx\R^2&3&
\begin{array}{c}
\R^2\cr\R\cr 
\end{array}
&\begin{array}{c}
\R\cr
\R^2 \cr
\end{array}\cr
\hline
\end{array}\cr
\end{eq}}

\noindent In addition, there are normal subgroups with  discrete factors
$\Z$ as used for $\D(1)\cong\R$.

There is Ado's theorem\cite{LIE13}: A finite dimensional Lie algebra
has a
faithful finite dimensional matrix re\-pre\-sen\-ta\-tions (the nilradical becomes 
strictly 
triangular), e.g. the Heisenberg Lie algebras above.

Any vector $v\in V$ of a group re\-pre\-sen\-ta\-tion space 
generates  - by the closure of its group orbit span 
(finite linear combinations) $\ol{\C^{(G\m v)}}$ -
a $G$-action invariant cyclic subspace.
 A vector $v$ is called 
cyclic for the representation if $\ol{\C^{(G\m v)}}=V$. A cyclic re\-pre\-sen\-ta\-tion has
 a cyclic vector $v$.
 Cyclic re\-pre\-sen\-ta\-tions have not to be irreducible (simple), e.g.
the reducible re\-pre\-sen\-ta\-tion\cite{BOE}
$\R\ni t\mape{\scriptsize\pmatrix{1&it\cr 0&1\cr}}\in\SU(1,1)$
with invariant subspace ${\scriptsize\pmatrix{\C\cr0\cr}}$
and cyclic vector ${\scriptsize\pmatrix{0\cr1\cr}}$.
This re\-pre\-sen\-ta\-tion describes the time development of a free Newtonian mass
point\cite{S89}
${\scriptsize\pmatrix{\bl x(t)\cr i\bl p(t)\cr}}=
{\scriptsize\pmatrix{1&-{it\over m}\cr 0&1\cr}} 
{\scriptsize\pmatrix{\bl x(0)\cr i\bl p(0)\cr}}$.
According to Maschke and Weyl\cite{MASCH,WEYLCG}, `irreducible' and `cyclic' 
coincide for compact groups.

Physical examples for cyclic vectors are  ground states
where a nontrivial fixgroup characterizes a degenerate ground state
(`sponteneous symmetry breakdown').
E.g., a ground state for the electroweak standard model
of quark and lepton fields with their interactions
is characterized by an electromagnetic $\U(1)$ as fixgroup (`little group')
in the represented interaction inducing hypercharge-isopin group $\U(2)$.

To define `realness' in a complex re\-pre\-sen\-ta\-tion of a real Lie group,
the re\-pre\-sen\-ta\-tion
vector space has to come with a conjugation,
 i.e. the represented Lie
group has to be a unitary group - definite or indefinite unitary.
E.g. the complex four dimensional Dirac re\-pre\-sen\-ta\-tion of
the real Lorentz group $\SL(\C^2)$ is a subgroup of the indefinite 
unitary group $\SU(2,2)$.

Group functions $\C^G=\{f:G\map\C\}$
are a `huge' re\-pre\-sen\-ta\-tion space of the `doubled' group $G\x G$ with
the both sided (left and right) regular action $f\stackrel{L_g\x R_k}\mape {_gf_k}$ where
$_gf_k(h)=f(g^{-1}hk)$. Complex group functions 
come  with the number induced conjugation
$f\lrmap\hat f$, $\hat f(g)=\ol{f(g^{-1})}$
(definite unitary $\U(1)$).

Of importance are
the Banach spaces with the Lebesque function classes $L^p(G)$, $1\le p\le\infty$,
 on a locally compact group\footnote{\scriptsize For
the function (classes)  $L^p_{d\mu}(S)$ where the $S$-measure $d\mu$ is unique
up to a scalar factor, e.g. Haar measure $dg$ for a locally compact group,
the measure is omitted in the notation $L^p(S)$.
Finite groups with discrete topology are compact 
and have counting measure.} with, especially, 
the Hilbert space with the square integrable functions
$L^2(G)$, the convolution group algebra $L^1(G)$ and, as its topological dual, 
the essentially bounded functions $L^\infty (G)$.
All Lebesque spaces are $L^1(G)$-convolution modules $L^1(G)*L^p(G)\map L^p(G)$.
For compact groups the group algebra is maximal $L^p(G)\sup L^q(G)$ if $p\le q$.

Representations in definite unitary groups are called 
Hilbert (re\-pre\-sen\-ta\-tions).
All compact group re\-pre\-sen\-ta\-tions are Hilbert.
Reducible 
re\-pre\-sen\-ta\-tions of compact groups are
decomposable into  
orthogonal direct sums of irreducible ones, 
the irreducible (cyclic, simple) re\-pre\-sen\-ta\-tions are 
- according to Weyl - finite dimensional.

In the twofold dichotomy abelian-nonabelian and compact-noncompact, exemplified
for finite dimensional real Lie groups with $r\ge1$

{\scriptsize\begin{eq}{c}
\begin{array}{|c||c|c|}\hline
&\hbox{abelian}&\hbox{nonabelian}\cr\hline\hline
\hbox{compact}&\U(1)&\SU(1+r)\cr\hline
\hbox{noncompact}&\D(1)&\SL(\C^{1+r})\cr\hline
\end{array}
\end{eq}}

\noindent the Hilbert re\-pre\-sen\-ta\-tion structure
of nonabelian noncompact
Lie operations, is much more complicated than that for compact and abelian ones.

According to Gelfand and Raikov\cite{GELRAI}, a Hilbert re\-pre\-sen\-ta\-tion of a locally compact group $G$
is an orthogonal  direct sum of cyclic ones and
- relevant for noncompact groups - an orthogonal  direct integral of irreducible ones.
Faithful Hilbert re\-pre\-sen\-ta\-tions
of noncompact locally compact groups are infinite dimensional.

Only for compact groups, all  representations act upon
Hilbert spaces with  square integrable functions
as group algebra subspaces $L^2(G)\sub L^1(G)$.
In general, the cyclic Hilbert 
re\-pre\-sen\-ta\-tions of a locally compact
group $G$ are - up to equivalence - bijectively related to  
positive type group functions. Such a function is defined
as an essentially bounded function 
$L^\infty(G)$
 which endows the group algebra with a definite product 
- $L^1(G)$ becomes a  pre-Hilbert space 
 \begin{eq}{l}
\left. \begin{array}{c}
 \om\in L^\infty(G)\cr
f\in L^1(G)\cr\end{array}\right\}
 \begin{array}{rl}\hbox{with }
 \sprod { f}{ f} _\om&=(\hat f*\om *f)(e)\cr
 &=\int_{G\x G} d\mu(g)d\mu(g') 
\ol{ f(g^{-1})}\om(gg') f(g')\ge0\end{array}

\end{eq}Any vector of a Hilbert re\-pre\-sen\-ta\-tion space
gives, by its diagonal matrix elements, a positive type function $G\ni g\mape
 \sprod v {g\m v}=\om(g)$.
 The positive type functions are diagonal matrix elements
 of  cyclic vectors.

Irreducible re\-pre\-sen\-ta\-tions 
of locally compact groups are characterized by invariants, constituing
the dual group space, for Hilbert re\-pre\-sen\-ta\-tions 
the definite dual group space.
The definite dual group space (invariants) comes with a Plancherel measure,
uniquely associated to a Haar measure of the group.
Locally compact noncompact nonabelian groups have 
also Hilbert re\-pre\-sen\-ta\-tions with
trivial Plancherel measure,
e.g. the supplementary re\-pre\-sen\-ta\-tions of the
nonabelian Lorentz groups $\SO_0(1,s)$, $s\ge2$.

\subsection
[Hilbert Representations of  Abelian and Compact Groups]
{Hilbert Representations\\of  Abelian and Compact
 Groups}

The Hilbert re\-pre\-sen\-ta\-tions of the abelian Lie subgroups
$\D(1)\cong\R$ and $\U(1)$ are basic for Hilbert re\-pre\-sen\-ta\-tions of all real Lie
groups.

The irreducible Hilbert re\-pre\-sen\-ta\-tions of 
the abelian noncompact groups (`translations' of rank $r$)
are $\R^r\ni x\mape e^{ipx}\in\U(1)$
(unfaithful).
The 1-di\-men\-sio\-nal Hilbert spaces $\C\rstate p$ are spanned by 
one normalized  eigenvector $\sprod pp=1$
with the translation behavior $\rstate {p(x)}=e^{ipx}\rstate p$
and the matrix element $\sprod {p(y)}{p(x)}=e^{ip(x-y)}$. 
In physics, the  eigenvalues $p\in\R^r$ as linear invariants are used as energy-mo\-men\-ta. 
The  combinations $\{\cos px, \sin px\}$ are
 matrix elements of selfdual re\-pre\-sen\-ta\-tions.
There, the Hilbert space $\C\rstate p\pl \C\lstate p$
 is spanned by the dual basic vectors
of the irreducible re\-pre\-sen\-ta\-tions.

In physics, translation re\-pre\-sen\-ta\-tions characterize free states, e.g. free
scattering states in position space $\R^3$ or free particles in spacetime $\R^4$.
This structure is  familiar from the simplest example, 
the harmonic oscillator with frequency (energy) $E\in\R$  where the 
creation operator
gives the eigenvector $\ro u( E)\rstate 0=\rstate  E$ 
with the time translation action $\rstate { E(t)}=e^{i E t}\rstate  E$.
The position and mo\-men\-tum operator 
 $\bl x={\ro u( E)+\ro u^\star ( E)\over \sqrt2}$ 
and  $i\bl p={\ro u( E)-\ro u^\star ( E)\over \sqrt2}$
are linear combinations of
creation and annihilation  operators and 
span the selfdual re\-pre\-sen\-ta\-tion space
with the time development of position-mo\-men\-tum
$\R\ni t\mape 
{\scriptsize\pmatrix{\cos E t&i\sin E t\cr i\sin E t&\cos E t\cr}} \in\SO(2)$
involving $\sprod{\bl x(s)}{\bl x(t)}=i\sin E(t-s)$ etc.

 The Plancherel measure,
associated to the Haar-Lebesque measure $d^rx$  of the translations
$\R^r$, is the Haar-Lebesque  measure  $d^r{p\over2\pi}$ of the 
(energy-)mo\-men\-ta $\R^r$ (dual group with linear invariants).
It comes with Schur's orthogonality\cite{FOL} for the
matrix elements of inequivalent re\-pre\-sen\-ta\-tions
by integrating with a Haar measure over the group
\begin{eq}{l}
\int d^rx~e^{ipx}e^{-ip'x}=\de({p-p'\over2\pi})
\end{eq}and is used for
the harmonic analysis of the translation functions
 $L^2(\R^r)$
(Fourier integrals).

If  functions, acted upon with a representation of 
a space and time translation group $x\in\R^r,$
are Fourier transformable $(f,\om)(x)=\int{d^rp\over(2\pi)^r}~
(\tilde f,\tilde\om)(p)e^{-ipx}$, a positive type function $\om$ for the scalar
product
\begin{eq}{l}
\sprod{f_1}{f_2}_\om
=\int d^nx~d^nx'~\ol{f_1(x)}~\om(x'-x)~f_2(x)
=\int {d^rp\over(2\pi)^r}~\ol{\tilde f_1(p)}~\tilde\om(p)~\tilde f_2(p)\cr
\end{eq}is expressed with a positive distribution
 $\tilde\om$ for the dual group with energies and momenta $p\in\R^r$.

The irreducible Hilbert re\-pre\-sen\-ta\-tions $e^{i\phi}\mape e^{iz\phi}$ of 
the compact quotient  $\R/\Z\cong\U(1)\cong\SO(2)$,
faithful for $z\ne0$, are given with the 
winding numbers (linear invariants) 
constituing the discrete  dual group
$z\in \Z\cong\R/\U(1)$. A physical example are 
the electromagnetic charge numbers,
integer multiples of one basic charge.  
 Again, the 1-di\-men\-sio\-nal Hilbert spaces $\C\rstate z$  are spanned by
one normalized eigenvector  $\sprod zz=1$.  
The Plancherel measure
associated to the normalized Haar measure $d{\phi\over2\pi}$ 
 of $\U(1)$ is the counting measure (dimension)
 $d(z)=1$ of the 
winding numbers (dual group space).
Schur's orthogonality reads 
 $\int_0^{2\pi}d{\phi\over2\pi}e^{iz\phi}e^{-iz'\phi}=\de_{zz'}$.
  It is used for 
the harmonic analysis of the group functions $L^p(\U(1))$
(Fourier series).

Compact groups of rank $r$ have discrete eigenvalues (weights)
in $\Z^r$, the dual groups of  $\U(1)^r$, e.g. (iso)spin $\SU(2)$ or 
color $\SU(3)$ or flavor groups $\SU(1+r)$. 
Extending $\U(1)$ by spherical degrees of freedom, the irreducible Hilbert re\-pre\-sen\-ta\-tions of 
(iso)spin $\SU(2)\ni u\mape 2J(u)$ and its quotient $\SO(3)$, e.g. 
for $J=1$ with Euler angles 
\begin{eq}{rl}
\SU(2)\ni u&=
{\scriptsize\pmatrix{
e^{i{\phi+\chi\over2}}\cos{\th\over2}
&ie^{-i{\phi-\chi\over2}}\sin{\th\over2}\cr
ie^{i{\phi-\chi\over2}}\sin{\th\over2}&
e^{-i{\phi+\chi\over2}}\cos{\th\over2}\cr}}\cr\cr
\mape 2(u)&=
{\scriptsize\pmatrix{
e^{i(\phi+\chi)}\cos^2{\th\over2}&
ie^{i\chi}{\sin \th\over\sqrt2}&
-e^{-i(\phi-\chi)}\sin^2{\th\over2}\cr
ie^{i\phi}{\sin\th\over\sqrt2}&
\cos \th&
ie^{-i\phi}{\sin\th\over\sqrt2}\cr
-e^{i(\phi-\chi)}\sin^2{\th\over2}&
ie^{-i\chi}{\sin \th\over\sqrt2}&
e^{-i(\phi+\chi)}\cos^2{\th\over2}\cr}}\in\SO(3)\subnoteq\SU(3)\cr
\end{eq}are given with the invariants 
$2J\in\N$ (spins as dual group space)
with scalar  product $\sprod {J;a'}{J;a}=\de^a_{-a'}$
(for  spherical bases).
The Plancherel counting  measure
$d(J)=1+2J$ (dimension of the irreducible Hilbert spaces)
associated to the normalized Haar measure $d^3u$  of $\SU(2)$
can be read off Schur's orthogonality
for the re\-pre\-sen\-ta\-tion matrix elements 
\begin{eq}{rl}
&\int _{\SU(2)}d^3u~~\ol{2J'(u)_{b'}^{a'}}{2J(u)_{b}^{a}}=
{1\over1+2J}\de_{JJ'}\de_{-a'}^{a}\de^{-b'}_b\cr
\hbox{e.g. for $J=1$:}&
\int_{-2\pi}^{2\pi}d{\chi\over4\pi}
\int_{0}^{2\pi}d{\phi\over2\pi}
\int_{-1}^{1}d{\cos\th\over2}|e^{i(\phi+\chi)}\sin^2{\th\over2}|^2={1\over3}

\end{eq}The Plancherel measure 
$\SUM_{2J=0}^\infty(1+2J)$ is used in the harmonic expansion of spin
group functions $L^2(\SU(2))$ as an example for the 
Peter-Weyl theorem\cite{PWEYL} for compact groups.

\subsection{Induced Representations}

All Hilbert re\-pre\-sen\-ta\-tions
of a locally compact Lie  group $G$ are inducable\footnote{\scriptsize
The basic structure for induced group actions is
the group left  action on  subgroup  right classes
$G\x G/H\map G/H$, $(k,gH)\map kgH$.}
 from those of its 
closed subgroups $\{H\sub G\}$.
The induced re\-pre\-sen\-ta\-tions\cite{FRO,WIG,MACK1,MACKP,FOL} act upon the vector space with the 
 mappings $W^{G/H}$ 
from the $H$-classes of the group  into a vector space $W$ with a
 Hilbert re\-pre\-sen\-ta\-tion of the subgroup $H$
\begin{eq}{l}
w:G/H\map W,~~gH\mape w(gH)\cr
\end{eq}As discussed in the literature, the mappings $W^{G/H}$ 
have to be `appropriately' defined
with respect to `smoothness' and measurability.

For finite dimension $W\cong\C^d$ the mappings  can be expanded 
(decomposed) into a direct integral
with a $G$-invariant   measure $dgH$ for the classes, 
a $W$-basis $\{e^a(gH)\}_{1=1,\dots,d}$ for each coset $gH\in G/H$
and the function values as coefficients
\begin{eq}{l}
W^{G/H}\ni w=\plint d gH~ w(gH)_a e(gH)^a,~~\dim_\C W^{G/H}=d\card G/H\cr 
\end{eq}also in bra-ket-notation, e.g. $\rstate w$ and
$\rstate{gH,a}$, with the identity decomposition
(sum over $a$) 
\begin{eq}{l}
\id_{W^{G/H}}\cong\plint d gH~ ~\rstate{gH,a}\lstate{gH,a}
\end{eq}A positive measure has an associated Dirac distribution
which, for an orthonormal $W$-basis, defines a scalar product distribution
\begin{eq}{l}
\sprod{g'H,a'}{gH,a}=\de^a_{a'}\de(gH,g'H)\hbox{ where }
\int d gH ~\de(gH,g'H)f(g'H)=f(gH)\cr
\end{eq}

In general, the induced $G$-re\-pre\-sen\-ta\-tions
\begin{eq}{l}
G\x W^{G/H}\map W^{G/H},~~(k,w)\mape k\m\rstate w=\plint d gH~ w(kgH)_a
\rstate{kgH,a}
\end{eq}are highly decomposable, e.g. with Frobenius' reciprocity
theorem\cite{FRO} for
compact groups.

For practical purposes, convenient  subgroups $H\sub G$ have to be chosen.
E.g., the trivial subgroup $\{e\}\sub G$ induces the left regular 
$G$-re\-pre\-sen\-ta\-tion
on the complex group functions $\C^G$ which
contains all $G$-re\-pre\-sen\-ta\-tions.
A re\-pre\-sen\-ta\-tion of the  full group `induces itself' with $W^{G/G}\cong W$.

\subsection{Residual Representations}

The eigenvalues (weights) for eigenvectors in a Lie group re\-pre\-sen\-ta\-tion 
are from a discrete or continuous spectrum. They are linear 
Lie algebra forms, e.g. 
the winding numbers $z\in\Z\subnoteq \R$ for $\U(1)$, 
the spin directions $2J_3\in\Z\subnoteq\R$ for $\SU(2)$ or
the energies $E\in\R$ for time translations $\R$
 and the mo\-men\-ta $\rvec p\in\R^3$ for position trnaslations $\R^3$. The
invariants arise from linear Lie algebra forms
in the abelian case and from at least bilinear form in the nonabelian case, e.g. 
 the spin Casimir $\rvec J^2$ from the Killing form  or the Euclidean invariant $\rvec p^2$.
A group acts on its Lie algebra and its forms in the adjoint
and coadjoint re\-pre\-sen\-ta\-tion.
 There exist 
  formulations for  re\-pre\-sen\-ta\-tion matrix elements
as residues of functions on complex Lie algebra 
forms\cite{S96,S97,S002,S031}, e.g. for the abelian groups
\begin{eq}{l}
\U(1)\ni e^{imx}=
\oint{dp\over 2i\pi}{1\over p-m}e^{ipx}=
\int dp~\de(p-m)e^{ipx},~~m\in \R
\end{eq}A not so trivial example
is the residual
 re\-pre\-sen\-ta\-tion of the  matrix elements  of the simple group $\SU(2)$,
 involving the derived Dirac distribution, i.e. the derived  $2$-sphere
 measure, supported by the value for the invariant $\rvec p^2=n^2$ 
  \begin{eq}{l}
\SU(2)\ni e^{in\rvec x}=
\int {d^3p\over\pi}
(n\bl 1_2+\rvec p)\de'(n^2-\rvec p^2)e^{i\rvec p\rvec x}
=\bl 1_2\cos n| \rvec x|+i\rvec x{\sin n|\rvec x|\over |\rvec x|}\cr
\hskip20mm n=2J=0,1,\dots
\end{eq}Euclidean  $\R^n$-vectors
 are written with arrows, e.g. here with Pauli matrices
$\rvec x=x_a\si_a=
{\scriptsize\pmatrix{x_3&x_1-ix_2\cr x_1+ix_2&-x_3\cr}}\in\R^3$
(more below).

The Dirac and principal value  distributions are
imaginary and real part 
in the associated complex distribution 
\begin{eq}{rl}
a\in\R:&
{\Ga(1+N)\over (a\mp io)^{1+N}}=
{\Ga(1+N)\over a_\ro P^{1+N}}
\pm i\pi \de^{(N)}(-a) 
,~N=0,1,\dots\cr
\end{eq}(Derived) Dirac distributions will be also called (multi)pole distributions.

The  distributions of Lie algebra forms 
show, on the one side, the re\-pre\-sen\-ta\-tion characteristic 
invariants as complex singularities, e.g. as 
poles or as  support of distributions,
and, on the other side, the structure of the 
Hilbert space where the re\-pre\-sen\-ta\-tion
acts upon. The  measures and distributions
 lead to distributions of Hilbert bases,
to distributions of the scalar product
and to functions on the Lie algebra forms
as Hilbert space vectors (more below).

\section{Simple Poles for Translations}

Hilbert re\-pre\-sen\-ta\-tions of  affine subgroups 
$G\sx \R^n$ with a homogeneous group $G\sub\GL(\R^n)$ acting on translations $\R^n$ 
- in the following for the semidirect `basic physical Lie operations'
\begin{eq}{l}
\D(1)\sx\R\hbox{ and }
\SO(2)\sx\R^2,~~\SO_0(1,1)\sx\R^2,~~\bl H(1)\cong\R\sx\R^2
\end{eq}are inducable\cite{WIG,MACK1} 
 from  Hilbert re\-pre\-sen\-ta\-tions 
of direct product 
subgroups $H\x \R^n$ which involve a  fixgroup $H\sub G$
 of the (energy-)mo\-men\-ta from the dual group $\R^n$.
An appropriate measure for the homogeneous space $G/H$
is constructed from $G$-orbits of the  (energy-)mo\-men\-ta.
The homogeneous group restricts and collects 
 the translation re\-pre\-sen\-ta\-tions
$x\mape e^{ipx}$  with the invariants.
 E.g. for the nonrelativistic scattering group $\SO(3)\sx\R^3$, 
 all translation characters $e^{i\rvec p\rvec x}$ are collected with 
the mo\-men\-tum square  
$\{\rvec p\in\R^3\mid \rvec p^2=P^2\}$.


\subsection
{The Affine Group in one Dimension - Causal Time}

In the group 
${\scriptsize\pmatrix{1&0\cr -t&e^\psi\cr}}\in \D(1)\sx\R$, interpreted as
dilations $\D(1)$ acting upon time translations  $\R\ni t\mape e^\psi t$,
 the energies are the eigenvalues 
for the time translations
 with the dual dilation action $\R\ni E\mape e^{-\psi} E$.

A Cartan subalgebra is spanned by the dilation generator with
adjoint re\-pre\-sen\-ta\-tion 
$\ad l^1={\scriptsize\pmatrix{0&0\cr0&1\cr}}$
- for the translations $\ad l^2={\scriptsize\pmatrix{0&0\cr -1&0\cr}}$.
The Killing form of the Lie algebra
$\tr\ad l^a\o\ad l^b\cong{\scriptsize\pmatrix{1&0\cr0&0\cr}}$
is degenerate, there is no nontrivial invariant. 

The   Hilbert re\-pre\-sen\-ta\-tions are inducable from the
$\D(1)$-action on the  energies.
For trivial energy $E=0$ with full fixgroup  $\D(1)$ and, therefore, 
 trivial $\D(1)$-orbit
there are  the 1-di\-men\-sio\-nal re\-pre\-sen\-ta\-tions
$\D(1)\ni  e^\psi\mape e^{i m\psi}\in\U(1)$ given above.  
These re\-pre\-sen\-ta\-tions are faithful only for $\D(1)\sx\R/\R\cong\D(1)$.

Nontrivial energies $E=\pm |E|\ne 0$, 
have trivial fixgroup (little group)
and, therefore,  $\D(1)$-isomorphic
orbits $\R_\pm=(0,\pm\infty)$ (either  positive or negative energies).
 They 
 lead to the two equivalence classes of faithful Hilbert re\-pre\-sen\-ta\-tions,
 induced from 
the $\U(1)$-re\-pre\-sen\-ta\-tion of time translations 
$\R\ni t\mape e^{iEt}$.
These re\-pre\-sen\-ta\-tions are orbit-integrated
with invariant energy-measure 
and  the characteristic functions for
positive and negative energies to give
as  matrix elements for the two irreducible re\-pre\-sen\-ta\-tions
\begin{eq}{l}
t\mape \pm i\int dE~\vth(\pm E)e^{iEt}={1\over t\mp io}
={1\over t_\ro P}\pm i\pi \de(t)\hbox{ where }\left\{\begin{array}{rl}
\vth(\pm E)&={1\pm \ep(E)\over2}\cr
\ep(E)&={E\over |E|}\cr\end{array}\right.
\end{eq}

Now, the Hilbert spaces
for causal time re\-pre\-sen\-ta\-tions: In contrast to the 1-di\-men\-sio\-nal irreducible   re\-pre\-sen\-ta\-tions
$\R\ni t\mape e^{iEt}\in\U(1)$
with only one eigenvalue and eigenvector $\rstate E$ with  $\sprod EE=1$,
the  $\D(1)\sx\R$ re\-pre\-sen\-ta\-tions  use either all 
positive or all negative energies. 
Here, the irreducible Hilbert spaces have infinite dimensions.
Basis distributions  (no Hilbert vectors) on the $\D(1)$-orbit for the two Hilbert spaces 
 are given 
with  an orthogonal and positive scalar product distribution
\begin{eq}{l}
\{\rstate {E_\pm}\mid E=\pm|E|\in\R\}
\hbox{ with }\left\{\begin{array}{rl}
\int {dE\over2\pi}~\vth(\pm
E)\rstate{E_\pm}\lstate{E_\pm}&=\id_{L^2(\R_\pm}\cr
\rstate{E_\pm}&\stackrel{\R}\mape e^{iE_\pm t}\rstate{E_\pm}\cr
\sprod{E'_\pm}{E_\pm }&=\vth(\pm E)\de({E-E'\over2\pi})\cr
\sprod{E'_+}{E_- }&=0\end{array}\right.
\end{eq}Vectors $\rstate{f_\pm}$  from the Hilbert spaces with their scalar product use
energy packets $E\mape  f(E)$
\begin{eq}{rl}
\rstate{f_\pm}=\int {dE\over2\pi}~f(E)\rstate{E_\pm}\then
\sprod{f'_+}{f_+}&=
\int_0^\infty  {dE\over2\pi} ~\ol{f'_+(E)}f_+(E)\cr 
\end{eq}The two types of re\-pre\-sen\-ta\-tions
act upon functions $f_\pm\in L^2(\R_\pm)$, square integrable on the 
$\D(1)$-energy orbits, i.e. on the  positive  and on the 
negative energies. The Hilbert product, written with time dependent functions,
employs the advanced and retarded distribution
 \begin{eq}{rl}
f_\pm(E)&=
\vth(\pm E) \int dt~  \tilde f(t)~e^{iEt}\cr
\then \sprod{f'_+}{f_+}&=
\int dt dt'~\ol{\tilde f'(t')}{1\over 2i\pi}{1\over t-t'-io}
\tilde f(t)\cr 

\end{eq}The translation function scalar product
$\int dt dt'~\ol{\tilde f'(t')}\de(t-t')\tilde f(t)$
is restricted corresponding to the action of the
homogeneous dilation group $\D(1)$.  

The dilation invariance of the  scalar product distribution 
determines  the dilation behavior of the basis distribution
\begin{eq}{rl}
E\mape e^{-\psi}E&\then\de(E-E')\mape e^\psi\de(E-E')\cr
&\then \rstate{E_\pm}\mape e^{\psi\over2}\rstate{E_\pm}
\end{eq}


\subsection
[The Flat Euclidean  Group - Scattering in the Plane]
{The Flat Euclidean  Group \\ - Scattering in the Plane}

In the flat Euclidean group  
nontrivial mo\-men\-ta 
have trivial fixgroup
\begin{eq}{l}
\SO(2)\sx\R^2\map\R^2:~
{\scriptsize\pmatrix{
\cos\th&-\sin\th\cr
\sin\th&\cos\th\cr}}
{\scriptsize\pmatrix{p_1\cr p_2\cr}}
={\scriptsize\pmatrix{p_1\cr p_2\cr}}\then H_{p\ne0}=\{1\}\cr
\end{eq}Their orbits  are isomorphic to the homogeneous group
$\SO(2)$, i.e. to circles in the mo\-men\-tum plane.
Therewith, the Hilbert spaces with faithful re\-pre\-sen\-ta\-tions  
are  orthogonal direct integrals 
over the 1-sphere 
with the normalized Haar measure 
\begin{eq}{rl}
\begin{array}{c}
\hbox{for }\SO(2)\cr
\hbox{with }P>0\end{array}:&\left\{\begin{array}{rl}	 
\int {d^2p\over \pi}~\de(\rvec p^2-P^2)&=\int _{-P}^P 
 {dp_1\over \pi \sqrt{P^2-p_1^2}}=\int_0^{2\pi} {d\th\over
2\pi}\cr
\rvec p^2=p_1^2+p_2^2,&{p_2\over p_1}=\tan\th\end{array}\right.\cr
\end{eq}

The translation re\-pre\-sen\-ta\-tions 
$\R^2\ni \rvec x\mape e^{i\rvec p\rvec x}$
have the mo\-men\-ta as eigenvalues with a positive invariant 
mo\-men\-tum squared  $P^2$
 for the Casimir element
characterizing an irreducible re\-pre\-sen\-ta\-tion. 
The  matrix elements
for the irreducible re\-pre\-sen\-ta\-tions
\begin{eq}{rl}
P^2>0:~\rvec x\mape&
\int {d^2p\over \pi}~\de(\rvec p^2-P^2)e^{i\rvec p\rvec x}
=\cl J_0( P|\rvec x|)
\end{eq}contain the integer index Bessel function\cite{VIL}
 (appendix `Residual Distributions')
\begin{eq}{l}
\R\ni \xi\mape  \cl J_0(\xi)=
\int_0^{\pi}{d\th\over\pi}~\cos(\xi\cos\th)=
{\SUM_{k=0}^\infty}{(-{\xi^2\over 4})^k\over
(k!)^2},~~
\cl J_0(0)=1
\end{eq}

By derivatives ${\p\over\p \rvec x}$ one obtains  matrix elements 
for nontrivial re\-pre\-sen\-ta\-tions of  axial rotations
$\SO(2)$.

The infinite dimensional Hilbert spaces with the irreducible re\-pre\-sen\-ta\-tions  
for $P^2>0$ have a 
 basis distribution of scattering `states'
in the plane $\R^2$  with 
the mo\-men\-ta on the $\SO(2)$-orbit, i.e. on 
the circle  with 
fixed mo\-men\-tum radius $P$. They have the  orthogonal and positive
distribution of the scalar product
\begin{eq}{l}
\{\rstate{P^2;\th}\mid 0\le\th<2\pi\}
\hbox{ with }\left\{\begin{array}{rl}
\int{d\th\over2\pi}\rstate{P^2;\th}\lstate{P^2;\th}&\cong\id_{L^2(\SO(2))}\cr
\rstate{P^2;\th}&\stackrel{\R^2}\mape e^{i\rvec p\rvec x}\rstate{P^2;\th}\cr
\hbox{with }\rvec p&=P(\cos\th,\sin\th)\cr
\sprod{P^2;\th'}{P^2;\th} &=\de({\th-\th'\over2\pi})\cr
\end{array}\right.
\end{eq}The Hilbert space vectors $\rstate{P^2;f}$ are
square integrable wave packets $f\in L^2(\Om^1)$
on the mo\-men\-tum sphere
\begin{eq}{rll}
\rstate{P^2;f}&=\int_0^{2\pi} {d\th\over2\pi}~f(\th)\rstate{P^2;\th}
&=\int {d^2p\over\pi}~\de(\rvec p^2-P^2)f(\rvec p)\rstate{\rvec p}\cr
\sprod{P^2;f'}{P^2;f}
&=\int_0^{2\pi} {d\th\over2\pi}~\ol{f'(\th)}f(\th)
&=\int {d^2p\over \pi}~\ol{f'(\rvec p)}\de(\rvec p^2-P^2)f(\rvec p)\cr
\end{eq}The Hilbert product in translation dependent functions
$f(\rvec p) = \int d^2x ~\tilde f(\rvec x) 
 e^{i\rvec p\rvec x}$
employs the Bessel function which modifies the scalar product
 for translation functions 
$\int {d^2xd^2x'\over2}~\ol{\tilde f'(\rvec x')}\de(\rvec x-\rvec x')
 \tilde f(\rvec x)$ according to the action of the
homogeneous group $\SO(2)$
\begin{eq}{rl}
\sprod{P^2;f'}{P^2;f}
&=\int {d^2xd^2x'\over2}~\ol{\tilde f'(\rvec x')}\cl J_0(P|\rvec x-\rvec x'|)
\tilde f(\rvec x)\cr
\end{eq}

The angle, momentum und translation dependent 
functions could be written
 with the same symbol 
 $(f(\th),f(\rvec p),f(\rvec x))$
  - the same function expanded in $\th$, $\rvec p$ or in $\rvec x$.
 Somewhat inconseqently, the notation is   
$(f(\th),f(\rvec p),\tilde f(\rvec x))$.

Schur's orthogonality for the square integrable re\-pre\-sen\-ta\-tion matrix elements 
involves the integration over the translations, e.g.
\begin{eq}{rl}
\int d^2x~\cl J_0(P|\rvec x|)\cl J_0(P'|\rvec x|)
&=4\pi\de(P^2-P'{}^2)\cr
\end{eq}This replaces, in 2-dimensional position 
space (in general in even dimensional
position), the Huygen-Fresnel principle with spherical Bessel functions,
e.g. 
$j_0(P|\rvec x|)={\sin P|\rvec x|\over P|\rvec x|} 
=\int{d^3 p\over2\pi}~{1\over |P| }\de(\rvec p^2-P^2)e^{i\rvec p\rvec x}$
 of 3-dimensional
(in general odd dimensional) position $\SO(3)\sx\R^3$.


\subsection
[The Flat  Poincar\'e Group - Spinless Free Particles]
{The Flat Poincar\'e Group\\
 - Spinless Free Particles}

Also in the flat  Poincar\'e group  
nontrivial  energy-mo\-men\-ta
have trivial fixgroup
\begin{eq}{l}
\SO_0(1,1)\sx\R^2\map\R^2:~
{\scriptsize\pmatrix{
\cosh\psi&-\sinh\psi\cr
-\sinh\psi&\cosh\psi\cr}}
{\scriptsize\pmatrix{q_0\cr q_1\cr}}
={\scriptsize\pmatrix{q_0\cr q_1\cr}}\then H_{q\ne0}=\{1\}\cr
\end{eq}The orbits  are - up to for the
lightcone energy-mo\-men\-ta $q^2=0$  - isomorphic to hyperbolas
$\SO_0(1,1)\cong\cl Y^1$ in the energy-mo\-men\-tum plane. 
Therewith, the Hilbert spaces with faithful re\-pre\-sen\-ta\-tions  
are  orthogonal direct integrals 
with Haar measure 
\begin{eq}{rl}
\hbox{for }\SO_0(1,1):&\left\{\begin{array}{rl}
\int {d^2q\over\pi}~\vth(\pm q_0)\de(q^2-m^2)&=
\int_0 ^\infty  {dq_1\over 2\pi\sqrt{q_1^2+m^2}}=\int_0 ^\infty
{d\psi\over2\pi} \cr
q^2=q_0^2-q_1^2,&{q_1\over q_0}=\tanh\psi\end{array}\right.\cr

\end{eq}The measure can be pa\-ra\-me\-tri\-zed  with mo\-men\-ta
 as familiar  from
particle quantum fields in 4-di\-men\-sio\-nal Minkowski spacetime
or with a hyperbolic `angle'.
  
For 2-di\-men\-sio\-nal Minkowski spacetime, there is an obvious isomorphy
between spacelike $y^2<0$ and timelike $y^2>0$ with
a timelike and a spacelike order structure.

The translation re\-pre\-sen\-ta\-tion matrix elements
have the energy-mo\-men\-ta  as eigenvalues with 
positive and negative invariant (`mass') for the Casimir element
$Q^2$. The  matrix elements
for the two types of inequivalent  irreducible re\-pre\-sen\-ta\-tions
$\{\pm m^2\mid m^2>0\}$
with nontrivial invariant
\begin{eq}{rll}
m^2>0:~y\mape&
\int {d^2q\over\pi}\de(q^2\mp m^2)e^{iqy}
&=\vth(\mp y^2){2\over\pi}\cl K_0(|my|)-\vth(\pm y^2)\cl N_0(|my|)\cr
&\hfill\hbox{ with } |y|&=\sqrt{|y^2|}\ge0\cr
\end{eq}come with  
the order 0  Macdonald function
\begin{eq}{rl}
\R\ni \xi\mape
2\cl K_0(\xi)
&=\int d\psi~e^{-|\xi|\cosh\psi}
=-
{\SUM_{n=0}^\infty}{ ({\xi^2\over4})^n\over (n!)^2}
[\log{\xi^2\over4}-2\Ga'(1)-2\phi(n)]\cr
\phi(0)&=0,~~\phi(n)=1+{1\over2}+\ldots +{1\over n},~~n=1,2,\dots \cr
-\Ga'(1)&=\lim_{n\to\infty}\brack{\phi(n)-\log n}=0.5772\ldots\cr
\end{eq}which - for imaginary argument -
is the Neumann function $\cl N_0$ 
with the Bessel function $\cl J_0$ 
\begin{eq}{l}
2\cl K_0(i\xi)=\int d\psi~e^{i|\xi|\cosh\psi}
=-\pi\cl N_0(\xi)+i\pi\cl J_0(\xi)

\end{eq}

By derivatives ${\p\over\p x}$ one obtains 
re\-pre\-sen\-ta\-tion matrix elements 
with nontrivial boost (dilation) $\SO_0(1,1)$-properties.

The re\-pre\-sen\-ta\-tion matrix elements for positive translation invariant
are familiar as the on-shell part of the Feynman propagator 
for massive particles
\begin{eq}{l}
{1\over i\pi}{1\over q^2-io-m^2}=\de(q^2-m^2)+{1\over i\pi}{1\over q_\ro
P^2-m^2}
\end{eq}Only the on-shell part gives
a Hilbert re\-pre\-sen\-ta\-tion matrix element of the Poincar\'e group $\SO_0(1,s)\sx\R^{1+s}$.

The Hilbert spaces with the irreducible re\-pre\-sen\-ta\-tions -
have an orthogonal and positive   basis distribution of free particles with 
energy-mo\-men\-ta $q= (\sqrt {m^2+q_1^2},q_1)$ on the 
 two shell hyperbola of mass $\pm m^2\ne0$
\begin{eq}{rll}
\{\rstate{m^2 ;q_1}\mid q_1\in\R\}
&\hbox{with }\sprod{m^2;q_1'}{m^2;q_1}&=
\sqrt{m^2+q_1^2}~\de({q_1-q_1'\over2\pi})\cr
\{\rstate{-m^2 ;q_0}\mid q_0\in\R\}
&\hbox{with }\sprod{-m^2;q_0'}{-m^2;q_0}&=
\sqrt{m^2+q_0^2}~\de({q_0-q_0'\over2\pi})\cr
\end{eq}

From now on, explicitly  only for
\begin{eq}{l}
\{\rstate{m^2;\psi}\mid \psi\in\R\}\hbox{ with }\left\{\begin{array}{rl}
\int {d\psi\over2\pi}\rstate{m^2;\psi}\lstate{m^2;\psi}&=
\id_{L^2(\SO(1,1))}\cr
\rstate{m^2;\psi}&\stackrel{\R^2}\mape e^{iqy}\rstate{m^2;\psi}\cr
\hbox{with }q&=m(\pm\cosh\psi,\sinh\psi)\cr
\sprod{m^2;\psi'}{m^2;\psi}&=
\de({\psi-\psi'\over2\pi})\cr

\end{array}\right.
\end{eq}The square integrable Hilbert space vectors are
wave packets $f\in L^2(\SO(1,1))$
of mo\-men\-ta on  the mass hyperboloid $\cl Y_2^1=\cl Y^1_+\cup\cl
Y^1_-\cong\SO(1,1)$ 
\begin{eq}{rll}
\rstate{m^2;f}&=
\int {d\psi\over2\pi}
f(\psi)\rstate{m^2;\psi}
&=\int {d^2q\over\pi}\de(q^2-m^2)f(q)\rstate{q}\cr
 \sprod{m^2;f'}{m^2;f}&=
\int {d\psi\over2\pi}
\ol{f'(\psi)}f(\psi)&=\int {d^2q\over\pi}
\ol{f'(q)}\de(q^2-m^2)f(q)\cr

\end{eq}For spacetime translation dependent functions
$f(q) = \int d^2y ~\tilde f(y) e^{iqy}$, the Hilbert product is restricted
according to the homogeneous action with the orthochronous
Lorentz group by the change of  $\de(y-y')$ into
the  combination of Macdonald and Neumann function 
\begin{eq}{rl}
\sprod{m^2;f'}{m^2;f}
=&\int {d^2yd^2y'\over2}~\ol{\tilde f'(y')}
\Bigl[\vth(- z^2){2\over\pi}\cl K_0(|mz|)
-\vth(z^2)\cl N_0(|mz|)
\Bigr]_{z=y-y'}
\tilde f(y)\cr
\end{eq}

If, for  Schur's orthogonality 
with  different invariants,
the integration is performed over the
translations there remains the infinite measure of the hyperboloid
\begin{eq}{rl}
\int d^2y\int {d^2q\over\pi}~\de(q^2-m^2)e^{iqy}\int {d^2q'\over\pi}~
\de(q'{}^2-m'{}^2)e^{iq'y}\cr
\hfill =\de(m^2-m'{}^2)
4\int d^2q~ \de( q^2-1)
\end{eq}The
  re\-pre\-sen\-ta\-tion matrix elements 
are not square integrable.

\subsection
[The Heisenberg Group - Nonrelativistic Quantum Mechanics]
{The Heisenberg Group \\ - Nonrelativistic Quantum Mechanics}

In the Heisenberg group 
as  semidirect  product $\bl H(1)=\R\sx\R^2$  
the homogeneous group $e^{q\bl x}\in\R$
with the position $\bl x\cong {\scriptsize\pmatrix{0&1\cr 0&0\cr}}$ 
acts on  the abelian normal subgroup $\R^2$ with 
mo\-men\-tum $\bl p\cong {\scriptsize\pmatrix{0\cr1\cr}}$
 and  the central action operator 
 $\bl I=[\bl x,\bl p]\cong {\scriptsize\pmatrix{1\cr0\cr}}$
\begin{eq}{l}
\R\sx\R^2\map\R^2:~~
{\scriptsize\pmatrix{1&q\cr 0&1\cr}}{\scriptsize\pmatrix{t\cr y\cr}}=
{\scriptsize\pmatrix{ t+qy\cr y\cr}}
\end{eq}$\bl I$ generates the  invariants.

The position action upon the dual space
$(\hbar ,p)\in\R^2$ with $(i\hbar ,ip)$ the eigenvalues 
for the action of $(\bl I,\bl p)$
\begin{eq}{l}
\dprod{(\hbar ,p)}{\scriptsize\pmatrix{t\cr y\cr}}=t\hbar +yp,~~

{(\hbar ,p)}{\scriptsize\pmatrix{1&-q\cr 0 &1\cr}}=
{(\hbar,p-\hbar q)}
\end{eq}has two types of  fixgroups 
with  corresponding orbits:
The fixgroups are characterized  either 
by trivial or by nontrivial eigenvalue   $\hbar\in\R$  
\begin{eq}{rlclll}
(\hbar =0,p)&\hbox{has full fixgroup}&\R&
\hbox{and  point orbit}&
\{(0,p)\}&\cong\{1\}\cr
{(\hbar \ne 0,p)}&\hbox{has trivial fixgroup}&\{1\}&
\hbox{and line orbit}&{(\hbar ,\R)}&\cong\R\cr
\end{eq}

Correspondingly, there are  two types of re\-pre\-sen\-ta\-tions
(Stone-von Neumann theorem\cite{NEUM}):
The  1st type with trivial
re\-pre\-sen\-ta\-tions of  the central action operator $\bl I\in\centr\bl H(1)$, i.e. 
invariant $\hbar =0$, leads
 to{ classical unfaithful re\-pre\-sen\-ta\-tions   of the  Heisenberg group}
with  commuting position and mo\-men\-tum,
i.e. of the 
abelian adjoint Heisenberg group  
$\Int\bl H(1)=\bl H(1)/\centr\bl H(1)\cong \R^2$.
The Hilbert re\-pre\-sen\-ta\-tions of  $\R^2$ are given above.
 
The 2nd type with trivial fixgroup
and a nontrivial   $\bl I$-eigenvalue  $i\hbar$ 
(there is a spectrum of $0\ne \hbar\in\R$) induces the quantum re\-pre\-sen\-ta\-tions  of the Heisenberg group.
Different action quanta $\hbar \ne \hbar'$ define inequivalent re\-pre\-sen\-ta\-tions
with $\bl I\mape i\hbar \bl 1$. 
These irreducible faithful re\-pre\-sen\-ta\-tions  integrate  
the irreducible  mo\-men\-tum $\bl p$-re\-pre\-sen\-ta\-tions 
$\R\ni y\mape e^{ipy}\in\U(1)$ for all
mo\-men\-ta eigenvalues on the orbit line $p\in \R$
with orthogonal and positive scalar product distribution
\begin{eq}{l}
\hbar \ne 0:~~\{\rstate{\hbar ;p}\mid p\in\R\}
\hbox{ with }\left\{\begin{array}{rl}
\int{dp\over2\pi}\rstate{\hbar ;p}\lstate{\hbar ;p}&\cong\id_{L^2(\R)}\cr
\rstate{\hbar;p}&\stackrel{\R}\mape e^{ipy}\rstate{\hbar;p}\cr
\sprod{\hbar ;p'}{\hbar ;p}&=\de({p-p'\over2\pi})\cr

\end{array}\right.
\end{eq}The Hilbert spaces consist 
of the square integrable
mo\-men\-tum functions $f\in L^2(\R)$
which are isomorphic to the square integrable
position functions 
$\tilde f(y)=\int dp~f(p)e^{ipy}$
\begin{eq}{rll}
\rstate{\hbar ;f}
&=\int {dp\over2\pi}~f(p)\rstate{\hbar ;p}\cr
\then\sprod{\hbar ;f'}{\hbar ;f}&=\int {dp\over2\pi}~\ol{f'(p)}f(p)
&=\int dy~\ol{\tilde f'(y)}\tilde f(y)\cr
\end{eq}The action of the Lie algebra mo\-men\-tum operator
is given by the derivative $\bl p\mape -i\hbar {d\over dy}$.

A harmonic analysis of   functions 
on the Heisenberg group $\bl H(n)$
uses  the classical Fourier components
  $\rstate{0;f}$ with trivial Plan\-che\-rel measure 
 and the quantum components $\rstate{\hbar ;f}$
 with Plan\-che\-rel measure\cite{FOL}
$|\hbar |^n d\hbar $ for the invariant values of  $\bl I$.

\section{Dipoles for Simple Groups}

In contrast to the re\-pre\-sen\-ta\-tions 
of abelian groups (translations) $\R^n$ with pole singularities
(Dirac distributions),
 simple groups use higher order poles (derived Dirac distributions).
This will be exemplified  for  
 the simplest simple compact and noncompact  Lie groups, the `basic Lie
 operations'  $\SU(2)$ and $\SU(1,1)$ with rank 1, 
 twofold covering $\SO(3)$ and $\SO_0(1,2)$, 
 the smallest nonabelian Lorentz group
for an odd dimensional spacetime 
$\SO_0(1,2R)$, $R=1,2,\dots$. Here
everything is explicitly known.
For the noncompact  group
$\SL(\R^2)\sim\SO_0(1,2)$, the definite  group dual has been given by 
Bargman\cite{BARG,GEL5}.

In the usual procedure (more detailed formulations in the literature)   
the group $ \SL(\R^2)\cong\SU(1,1)$ 
is used  in the defining re\-pre\-sen\-ta\-tion
on real 2-vectors and their component ratio
\begin{eq}{l}
{\scriptsize\pmatrix{\xi _1\cr \xi _2\cr}}
\mape {\scriptsize\pmatrix{a&b\cr c&d\cr}}
{\scriptsize\pmatrix{\xi _1\cr \xi _2\cr}}
={\scriptsize\pmatrix{a\xi _1+b\xi _2\cr c\xi _1+d\xi _2\cr}}\then
\xi = {\xi _1\over \xi _2}\mape {a\xi +b\over c\xi +d}\cr
\end{eq}The $\SL(\R^2)$-transformation behavior of complex functions of the ratio
$\xi \mape F(\xi )$, induced by a re\-pre\-sen\-ta\-tion of a Cartan subgroup 
$\SO(2)$ or $\SO_0(1,1)$ is given by  
\begin{eq}{l}
({\scriptsize\pmatrix{a&b\cr c&d\cr}}\m F)(\xi )=
 \eta_\mu(c\xi +d)^{-1-\mu}F({a\xi +b\over c\xi +d})\cr
\end{eq}where the power $\mu\in\C$ in the overall factor is the
Cartan subgroup  re\-pre\-sen\-ta\-tion characterizing invariant
and $\eta_\mu$ a re\-pre\-sen\-ta\-tion dependent sign factor.
The Cartan subgroup re\-pre\-sen\-ta\-tion
determines the homogeneouity property of the functions in one
irreducible re\-pre\-sen\-ta\-tion.

In the following,  an orientation for $\SU(2)$-Hilbert re\-pre\-sen\-ta\-tions 
- as a warm up - and for
$\SU(1,1)$-Hilbert re\-pre\-sen\-ta\-tions  is given by applying the inducing procedure
and  residual re\-pre\-sen\-ta\-tions in a spacetime and energy-mo\-men\-tum oriented language. 
For the noncompact simple group 
there arise both square integrable Hilbert spaces and 
Hilbert spaces defined with positive type functions.

\subsection{Multipoles for the Nonrelativistic Hydrogen Atom}
A familiar example of higher order poles for simple Lie operations,
discussed in more detail in\cite{S032}, are the Schr\"odinger bound states 
functions for
the nonrelativistic hydrogen atom with Hamiltonian 
${\rvec{\bl p}^2\over2}-{1\over |\rvec{\bl x}|}$. The related Lenz-Runge 
invariance (perihel conservation)\cite{FOCK} leads to the action group $\SO(4)$ with the rotation group classes
the 3-sphere $\Om^3\cong\SO(4)/\SO(3)$. The $\Om^3$-measure has 
a mo\-men\-tum pa\-ra\-me\-tri\-zation by dipoles 
\begin{eq}{l}
\R^4\ni {\scriptsize\pmatrix{
\cos\phi\cr
\rvec \om\sin\phi\cr}}={1\over\sqrt{\rvec p^2+1}}
{\scriptsize\pmatrix{
1\cr
\rvec p\cr}},~~
\int d\Om^3=\int d^3 p{2P\over (\rvec p^2+P^2)^2}=2\pi^2
\end{eq}with 
the imaginary `mo\-men\-tum' invariant $\rvec p^2=-P^2$.
Its Fourier transform is the hydrogen ground state 
$\int {d\Om^3\over 2\pi^2} ~e^{-i\rvec p\rvec x}=e^{-Pr}$
with binding energy $-2E=P^2=1$ for principal quantum number $k=1$.
It is a re\-pre\-sen\-ta\-tion matrix element of nonrelativistic position as
symmetric space
$\cl Y^1\x\Om^2\cong\R^3$. 
The $\Om^3$-spherical harmonics lead to higher order poles
for the $\SO(4)$-multiplets. E.g.,
 the position re\-pre\-sen\-ta\-tion matrix elements in
the quartet  $k=2$ (s- and p-states) come with normalized
$\SO(4)$-vectors 
\begin{eq}{l}
e(\rvec p)={1\over\rvec p^2+ 1}
{\scriptsize\left(\begin{array}{c}
\rvec p^2-1\cr
2i  \rvec p\cr
\end{array}\right)}\in\R^4,~~\sprod{e(\rvec p)}{e(\rvec p)}=1
\end{eq}leading to 3rd order poles
\begin{eq}{rl}
\int {d\Om^3\over 2\pi^2}~e({\rvec p\over P})e^{-i\rvec p\rvec x}&=
 \int{d^3p\over \pi^2}~
{P\over(\rvec p^2+ P^2)^3}
{\scriptsize\left(\begin{array}{c}
\rvec p^2-P^2\cr
2i P \rvec p\cr
\end{array}\right)}
e^{-i\rvec p\rvec x}={1\over2}
{\scriptsize\left(\begin{array}{c}
1-Pr\cr
P\rvec x\cr
\end{array}\right)}
e^{-Pr }\cr
\hbox{ with }  -2E&=P^2={1\over2^2} 
\end{eq}In general, there are  order $(1+k)$-poles 
with power $(k-1)$-tensors $e(\rvec p)^{k-1}$ for 
bound states with energy $-2E=P^2={1\over k^2}$ and $k=1,2,\dots$


\subsection{3-Space and (1,2)-Spacetime as Lie Algebras}

The groups $\SU(2)$ and $\SU(1,1)$ in the defining  representation
by $(2\x2)$-matrices $W\ox W^T\cong\C^2\ox\C^2$
have  a Lie algebra pa\-ra\-me\-tri\-zation
almost everywhere 
\begin{eq}{rl}
\SU(2)\ni  e^{i\rvec x }&=\bl1_2\cos|\rvec x|+{i\rvec x\over|\rvec
x|}\sin|\rvec x|\cr
\SU(1,1)\ni  e^{iy}&=
\vth(y^2)[\bl1_2\cos|y|+{iy\over|y|}\sin|y|]
\cr
&+\vth(-y^2)[\bl1_2\cosh|y|+{iy\over|y|}\sinh|y|]
\cr
\end{eq}using a 3-di\-men\-sio\-nal Euclidean space
$\rvec x\in i\log\SU(2)$  and $(1,2)$-spacetime $y\in i\log\SU(1,1)$ 
\begin{eq}{rll}
i\rvec x&=ix_3\si_3+ix_1\si_1+ix_2\si_2
&={\scriptsize\pmatrix{ix_3&ix_1+x_2\cr
ix_1-x_2&-ix_3}}\in\log\SU(2)\cong\R^{3}\cr
&\hfill\hbox{invariant }\rvec x^2&={1\over2}\tr\rvec x\o\rvec x=\det i\rvec x=x_3^2+x_1^2+x_2^2\cr
iy&
=iy_0\si_3+y_1\si_1+y_2\si_2
&={\scriptsize\pmatrix{iy_0&y_1-iy_2\cr
y_1+iy_2&-iy_0}}\in\log\SU(1,1)\cong\R^{1+2}\cr
&\hfill\hbox{invariant }y^2&={1\over2}\tr y\o y=\det iy= y_0^2-y_1^2-y_2^2\cr
&\hfill|y|&=\sqrt{|y^2|}
\end{eq}The invariants are 
 generated by the quadratic Killing invariant which gives the Euclidean
and Lorentz metric.
The compact parameter space is restricted to 
$\rvec x^2<(2\pi)^2$ for $\SU(2)$ and
to the timelike `subhyperboloid' $\vth(y^2)y^2<(2\pi)^2$ for $\SU(1,1)$.

The group acts adjointly on its Lie algebra
$\rvec x\mape u\o \rvec x\o u^{-1}$ for $u\in\SU(2)$ and
$y \mape v\o y\o v^{-1}$ for $v\in\SU(1,1)$.
Obviously in this case, 
the group transformations 
$u$ and $v$ cannot be pa\-ra\-me\-tri\-zed with $\rvec x$ and
$y$. E.g.  group pa\-ra\-me\-tri\-zations with independent Euler `angles'
can be used
\begin{eq}{rll}
u(\phi,\th,\chi)&=
{\scriptsize\pmatrix{
e^{i{\phi+\chi\over2}}\cos{\th\over2}&
ie^{i{\phi-\chi\over2}}\sin{\th\over2}\cr
ie^{-i{\phi-\chi\over2}}\sin{\th\over2}&
e^{-i{\phi+\chi\over2}}\cos{\th\over2}\cr}}
&=
e^{i{\phi\over2}\si_3}e^{i{\th\over2}\si_1}
e^{i{\chi\over2}\si_3}\in\SU(2)\cr\cr
v(\phi,\psi,\chi)&=
{\scriptsize\pmatrix{
e^{i{\phi+\chi\over2}}\cosh{\psi\over2}&
e^{i{\phi-\chi\over2}}\sinh{\psi\over2}\cr
e^{-i{\phi-\chi\over2}}\sinh{\psi\over2}&
e^{-i{\phi+\chi\over2}}\cosh{\psi\over2}\cr}}
&=
e^{i{\phi\over2}\si_3}e^{{\psi\over2}\si_1}
e^{i{\chi\over2}\si_3}
\in\SU(1,1)\cr
\end{eq}

The dual vector spaces, i.e. the linear forms  of the Lie algebras,
are pa\-ra\-me\-tri\-zed by mo\-men\-ta 
and by energy-mo\-men\-ta 
\begin{eq}{l}
\rvec p\in\R^3\cong(\log\SU(2))^T,~~
q\in\R^3\cong(\log\SU(1,1))^T
\end{eq}both with the corresponding coadjoint action.
The nontrivial $\SU(2)$-mo\-men\-ta, e.g. 
$\si_3$,
have a spherical  orbit pa\-ra\-me\-tri\-zation (polar coordinates)
where the  fixgroup $e^{i{\chi\over2}\si_3}\in\SO(2)$
parameter drops out
\begin{eq}{rl}
\SU(2):~~u(\phi,\th,\chi)\o \si_3\o u(\phi,\th,\chi)^{-1}
&={\scriptsize\pmatrix{
\cos\th&-ie^{i\phi}\sin\th\cr
ie^{-i\phi}\sin\th&-\cos\th\cr}}\cr
\rvec p^2>0\then 
{\scriptsize\pmatrix{p_3\cr p_1 \cr p_2\cr}}
&=\sqrt{\rvec p^2}{\scriptsize\pmatrix{\cos\th\cr\sin\th\sin\phi\cr \sin\th\cos\phi \cr }}
\cr
\end{eq}For the energylike $\SU(1,1)$-energy-mo\-men\-ta,
e.g. $\si_3$ with
fixgroup $e^{i{\chi\over2}\si_3}\in \SO(2)$, there are the corresponding  
hyperbolic 
orbit (`polar') coordinates 
 \begin{eq}{rl}
\SU(1,1):~~
v(\phi,\psi,\chi)\o \si_3 \o v(\phi,\psi,\chi)^{-1}
&={\scriptsize\pmatrix{
\cosh\psi&-e^{i\phi}\sinh\psi\cr
e^{-i\phi}\sinh\psi&-\cosh\psi\cr}}\cr
q^2>0\then {\scriptsize\pmatrix{q_0\cr q_1 \cr q_2\cr}}
&=\sqrt{ q^2}
{\scriptsize\pmatrix{\cosh\psi\cr \sinh\psi\sin\phi\cr \sinh\psi\cos\phi \cr }}
\cr
\end{eq}For the mo\-men\-tumlike energy-mo\-men\-ta, e.g. 
$i\si_1$ with
fixgroup $e^{{\xi\over2}\si_1}\in \SO_0(1,1)$,
another pa\-ra\-me\-tri\-zation with two noncompact parameters is appropriate
\begin{eq}{rl}
w(\phi,\psi,\xi)
&=e^{i{\phi\over2}\si_3}e^{{\psi\over2}\si_2}
e^{{\xi\over2}\si_1}\cr
&=e^{i{\phi\over2}\si_3}
{\scriptsize\pmatrix{
\cosh{\psi\over2}&-i\sinh{\psi\over2}\cr
i\sinh{\psi\over2}&\cosh{\psi\over2}\cr}}
{\scriptsize\pmatrix{
\cosh{\xi\over2}&\sinh{\xi\over2}\cr
\sinh{\xi\over2}&\cosh{\xi\over2}\cr}}
\in\SU(1,1)\cr
\end{eq}with  the hyperbolic orbit (`polar') coordinates
 \begin{eq}{rl}
\SU(1,1):~~
w(\phi,\psi,\xi)\o i\si_1\o w(\phi,\psi,\xi)^{-1}
&={\scriptsize\pmatrix{
\sinh\psi&ie^{i\phi}\cosh\psi\cr
ie^{-i\phi}\cosh\psi&-\sinh\psi\cr}}\cr
q^2<0\then {\scriptsize\pmatrix{q_0\cr q_1 \cr q_2\cr}}
&=\sqrt{- q^2}
{\scriptsize\pmatrix{\sinh\psi\cr 
-\cosh\psi\cos\phi\cr \cosh\psi\sin\phi \cr }}
\cr
\end{eq}


The semidirect groups $\SO(1+s)\sx\R^{1+s}$ and $\SO_0(1,s)\sx\R^{1+s}$
describe the adjoint action of the homogeneous groups on their Lie
algebras  for ${1+s\choose 2}=1+s$, i.e. only for $1+s =3$.
For $1+s=3$,  the Pauli spinor re\-pre\-sen\-ta\-tion is equivalent to
its conjugated one, i.e. for group  $g\cong\hat g=g^{-1\star}$ and for the
Lie algebra $l\cong \hat l=-l^\star$  
\begin{eq}{rl}
\hbox{for }\log\SU(2):&-(i\rvec x)^\star=i\rvec x\cr 
\hbox{for }\log\SU(1,1):&-(iy)^\star
=iy_0\si_3-y_1\si_1-y_2\si_2
=\si_3\o iy\o \si_3
\end{eq}

The orthogonal structures for $1+s=3$ are embedded into,
but not trivially generalizable to higher dimensions.
E.g., even dimensional
nonabelian orthogonal structures, starting with the rank 2 
 proper Lorentz group
$\SL(\C^2)\sim\SO_0(1,3)$, have two inequivalent left and right handed 
Weyl spinor
re\-pre\-sen\-ta\-tions, for the Lorentz group  $g=e^{i\rvec\al+\rvec\be}$ 
and $\hat g=e^{i\rvec\al-\rvec\be}$. 
The Hilbert representations of 
$\SL(\C^2)\sim\SO_0(1,3)$ have been given 
by Gel'fand and Naimark\cite{GELNAI,NAIM,GEL5}.

\subsection{Hilbert Representations of $\SU(2)$ and $\SU(1,1)$}

 The two types of maximal abelian subgroups 
(Cartan subgroups) in $\SU(1,1)$ and the corresponding
classes are pa\-ra\-me\-tri\-zable by time and position translations
\begin{eq}{ll}
y^2>0:~~e^{iy_0\si_3}\in \SO(2),&\SO_0(1,2)/\SO(2)\cong\cl Y^2\cr
 &\hbox{(1-shell `timelike' hyperboloid)}\cr
y^2<0:~~e^{y_1\si_1}\in \SO_0(1,1),&\SO_0(1,2)/\SO_0(1,1)\cong 
\cl Y^{(1,1)}=\cl Y^1\x\Om^1 \cr
&\hbox{(1-shell `spacelike' hyperboloid)}\cr
&\SO_0(1,2)\cong\SU(1,1)/\{\pm \bl1_2\}\end{eq}The Hilbert re\-pre\-sen\-ta\-tions
of the compact and noncompact Cartan subgroups have
discrete and continuous  eigenvalues $\{z,iq\}$ in their dual groups -
 the energy-mo\-men\-ta as linear Lie algebra forms
\begin{eq}{rll} 
\irrep\U(1)&\cong\Z
&\then 
~~e^{iy_0}\mape e^{izy_0}\in\U(1),~
 z\in\Z\cr
&&\hskip10mm\hbox{for discrete series}\cr
\irrep_+\D(1)&\cong i\R
&\then \left\{
\begin{array}{l}
e^{y_1}\mape e^{iq y_1}\in\U(1),~~
 iq\in i\R\cr
\hbox{for principal  series}\cr
\end{array}\right.
\end{eq}The quadratic Killing invariants $\mu^2\in\R$ 
are called polarization $n^2=z^2$ 
and mo\-men\-tumlike invariant 
$P^2=-q^2$.


The measures for  the compact classes $\SO(1+s)/\SO(s)\cong\Om^s$ (spheres)
and for the noncompact ones 
$\SO_0(1,s)/\SO(s)\cong\cl Y^s$ (energylike hyperboloids) and
$\SO_0(1,s)/\SO_0(1,s-1)\cong\cl Y^1\x\Om^{s-1}=\cl Y^{(1,s-1)}$ 
(mo\-men\-tumlike hyperboloids)
are 
\begin{eq}{rll}
\int d\Om^s&=\int d^{1+s}p~\de(\rvec p^2-1)&=
\int_0^\pi(\sin\th)^{s-1}d\th\int d\Om^{s-1}\cr
\int d\cl Y^s&=\int d^{1+s}q~\vth(\pm q_0)\de(q^2-1)&=
\int_0^\infty(\sinh\psi)^{s-1}d\psi\int d\Om^{s-1}\cr
\int d \cl Y^{(1,s-1)}&=\int d^{1+s}q~\de(q^2+1)&=\int_0^\infty
(\cosh\psi)^{s-1}d\psi\int d\Om^{s-1}\cr
\end{eq}

The distributions and residues for one  dimension
\begin{eq}{l}
\int {d q \over  i\pi}{1\over q^2-io- 1}e^{iqt}
=e^{i|t|},~~
\int {d q \over  \pi}{1\over q^2+1}e^{iqz}
=e^{-|z|}
\end{eq}are both  embedded in the two 3-di\-men\-sio\-nal distributions
with energylike and mo\-men\-tumlike invariant as poles 
whose  Dirac contributions are the
measures of $\Om^2$, $\cl Y_2^2$ and $\cl Y^{(1,1)}$
\begin{eq}{rl}
\SU(2):&\int {d^3 p \over  2\pi^2}{1\over \rvec p^2-io- 1}e^{i\rvec p\rvec x}
={e^{i|\rvec x|}\over |\rvec x|}\cr

\SU(1,1):&\left\{\begin{array}{rl}

\int {d^3 q \over  2\pi^2}{1\over q^2-io- 1}e^{iqy}
&={\vth(-y^2)ie^{-|y|}-\vth(y^2)e^{i|y|}\over|y|}\cr

\int {d^3 q \over  2\pi^2}{1\over
q^2-io+  1}e^{iqy}
&={-\vth(y^2)e^{-| y|}+i\vth(-y^2)e^{-i| y|}\over|y|}\cr\end{array}\right.
\end{eq}
 
The order structure of 
spacetime $\R^{1+s}$ is encoded  in the two causal distributions, 
the advanced and retarded one
\begin{eq}{rl}
{1\over (q\mp io)^2-1}&=
{1\over q_\ro
P^2-1}\pm i\pi\ep(q_0)\de(q^2-1)
\cr
\hbox{ with }
(q\mp io)^2&=(q_0\mp io)^2-\rvec q^2\cr
{1\over q^2\mp io-1}&=
{1\over q_\ro P^2-1}\pm i\pi\de(q^2-1)
\cr
\end{eq}The sum of advanced and retarded distribution  
coincides with the sum of the two conjugated Feynman distributions.
The Fourier transformation  of advanced and retarded distributions
are supported by  future and past 
\begin{eq}{rl}
\int {d^{1+s}q\over \pi} {1\over q_\ro P^2-1}e^{iqy}&
=i\ep(y_0)
\int d^{1+s}q~ \ep(q_0)\de(1-q^2)e^{iqy}\cr
\int{d^{1+s}q\over 2\pi}{1\over (q\mp io)^2-1}e^{iqy}&=
\pm i\vth(\pm y_0)\int d^{1+s}q~\ep(q_0)\de(1-q^2)e^{iqy}\cr
&=\vth(\pm y_0)\int {d^{1+s}q\over \pi} {1\over q_\ro P^2-1}e^{iqy}\cr

\end{eq}The Feynman distributions are 
compatible with the action of any orthogonal group $\O(p,q)$
with bilinear form $q^2$,  the causal distributions
only with  the action of the orthochronous Lorentz groups
$\SO_0(1,s)$.

One obtains for time and $(1,2)$-spacetime
\begin{eq}{rl}
\int{dq\over 2\pi}{1\over (q- io)^2-1}e^{iqt}&=
-\vth( t)\sin |t|\cr
\SU(1,1):~~\int{d^3q\over 4\pi^2}{1\over (q-  io)^2-1}e^{iqy}&=
-\vth(y^2)\vth( y_0){\cos |y|\over |y|}\cr
\end{eq}

The  $\SU(2)$-re\-pre\-sen\-ta\-tions  are   inducable 
from Cartan subgroup re\-pre\-sen\-ta\-tions
with winding numbers (integer powers)
$\SO(2)\ni e^{ix_3\si_3}\mape
(e^{ix_3\si_3})^{\pm n}\in\SO(2)$
and the decomposition $\SU(2)/\{\pm\bl1_2\}\cong\SO(3)\cong\SO(2)\x\Om^2$.
The  matrix elements for 
spin $J$ arise from  dipole distributions, supported by the invariant
\begin{eq}{l}
\begin{array}{r}
\SU(2)\ni e^{i\rvec x}\mape\cr 
\hbox{discrete } n=2J=1,2,\dots
\end{array}\left\{\begin{array}{rl}
\pm\int {d^3 p \over i\pi^2}{n\over (\rvec p^2\mp io-n^2)^2}e^{i\rvec p\rvec x}
&=e^{\pm i n|\rvec x|}\cr
\int{d^3 p\over\pi}~n~\de'(n^2-\rvec p^2)e^{i\rvec p\rvec x}&=
\cos n |\rvec x|\cr
\int{d^3 p\over \pi}~\rvec p~\de'(n^2-\rvec p^2)e^{i\rvec p\rvec x}&=
{i\rvec x\over|\rvec x|} |\sin n |\rvec x|\cr
\end{array}\right.
\end{eq}Matrix elements with nontrivial properties with respect
to the rotation classes $\SU(2)/\SO(2)$ use 
derivations ${\p\over\p\rvec x}={\rvec x\over |\rvec x|}{\p\over\p|\rvec x|}$.

For the $\SU(1,1)$-re\-pre\-sen\-ta\-tions,
there are six   dipole distributions (three pairs with $\pm io$).
They involve the Cartan subgroup re\-pre\-sen\-ta\-tion matrix elements 
$\{e^{i|y|},e^{-|y|}\}$
for the invariant $|y|=\sqrt{|y^2|}$.
There are two pairs  
of $\SU(1,1)$-re\-pre\-sen\-ta\-tion types: 
The  discrete `energylike' polarization  invariants 
give the two causally supported series, 
induced by compact Cartan subgroup representations
$\SO(2)\ni e^{iy_0\si_3}\mape
(e^{iy_0\si_3})^{\pm n}\in\SO(2)$
and the decomposition $\SU(1,1)/\{\pm\bl1_2\}\cong\SO_0(1,2)\cong\SO(2)\x\cl Y^2_\pm$
\begin{eq}{l}
\begin{array}{r}
\SU(1,1)\ni e^{iy}
\mape\cr
\hbox{discrete: }n=0,1,2,\dots\cr\end{array}
\left\{\begin{array}{ll}

n^2_\pm(y)
&=\int{d^3q\over 2\pi^2}~{n\over [(q\mp io)^2-n^2]^2}e^{iqy}\cr
&=\vth(y^2)\vth(\pm y_0)\sin n|y|\cr
n^2_\pm(y)'
&=\int{d^3q\over 2\pi^2}~{q\over [(q\mp io)^2-n^2]^2}e^{iqy}\cr
&=-\vth(y^2)\vth(\pm y_0){iy\over|y|}\cos n|y|\cr
\end{array}\right.
\end{eq}The lightlike invariant lead to 
the `mock' series $0_\pm$.
The two principal\footnote{\scriptsize
`Principal' in principal series and principal value are not related to each other}
 series are  induced by noncompact Cartan subgroup representations
$\SO_0(1,1)\ni e^{y_1\si_1}\mape
(e^{y_1\si_1})^{\pm i|P|}\in\SO(2)$
with imaginary eigenvalues and
a mo\-men\-tumlike invariant
and the decomposition $\SU(1,1)/\{\pm\bl1_2\}\cong\SO_0(1,2)\cong\SO_0(1,1)\x\cl Y^{(1,1)}$
\begin{eq}{l}
\begin{array}{r}
\SU(1,1)\ni e^{iy}\mape\cr
\hbox{principal: }P^2> 0\cr
\end{array}
\left\{\begin{array}{ll}
P^2_\pm(y)
&=
-\int {d^3 q \over  \pi^2}{|P|\over (q^2\mp io+  P^2)^2}e^{iqy}\cr
&=\vth(-y^2)e^{\mp i|Py|}+\vth(y^2)e^{-|Py|}\cr
\end{array}\right.

\end{eq}In addition, there is the
causally supported supplementary series 
with principal value integration 
 \begin{eq}{l}
\hfill\hbox{from } \mp\int {d^3 q \over  i\pi^2}{1\over (q^2\mp  io-1)^2}e^{iqy}
=\vth(y^2)e^{\pm i|y|}+\vth(-y^2)e^{-|y|}\cr
\begin{array}{r}
\SU(1,1)\ni e^{iy}\mape\cr
\hbox{supplementary: }0< m^2<1\end{array}
\left\{\begin{array}{ll}
m^2_{\ro P}(y)&=\int {d^3 q \over  \pi^2}{|m|\over (q_\ro P^2-m^2)^2}e^{iqy}\cr
&=\vth(y^2)\sin |my|\cr
m^2_{\ro P}(y)'&=\int {d^3 q \over  \pi^2}{q\over (q_\ro P^2-m^2)^2}e^{iqy}\cr
&=-\vth(y^2){iy\over|y|}\cos |my|\cr
\end{array}\right.
\end{eq}These representations  can be understood to be induced from 
a noncompact Cartan subgroup  representation  by
 a rotation of a spacelike direction into a `timelike' one
 $\si_1\mape \si_3$ (not $i\si_3$) and exponentiating  
with a continuous imaginary eigenvalue
\begin{eq}{rl}
\SO_0(1,1)\ni e^{y_1\si_1}&\mape
(w \o e^{y_1\si_1}\o w^*)^{\pm i|m|}
=(e^{y_1\si_3})^{\pm i|m|}
\in\SO(2)\cr
\hbox{with }w&=e^{i{\pi\over4}\si_2}\in\SU(2)
\end{eq}With the compact $\SU(1,1)$-parameter space, the continuous energylike 
invariant  is restricted by the lowest nontrivial polarization $n^2=1$.

Matrix elements with nontrivial properties with respect
to the axial rotation classes $\SU(1,1)/\SO(2)$ and boost classes 
  $\SU(1,1)/\SO_0(1,1)$ use
   derivations ${\p\over\p y}={y\over|y|}{\p\over\p |y|}$.

The Plancherel measure\cite{STRICH,FOL} for the supplementary and the trivial 
re\-pre\-sen\-ta\-tions is trivial, the discrete re\-pre\-sen\-ta\-tions, induced from $\SO(2)$,
have counting measure $\mu(n^2_\pm)=n$ and the principal ones, 
induced from $\SO_0(1,1)$, a hyperbolic measure
$d\mu(P_\pm^2)
=(\tanh {\pi P\over2},\coth {\pi P\over2}) dP^2$
(always with fixed Haar measure).

\subsection{Hilbert Spaces of $\SU(2)$}

The following derivation 
of  the familiar $\SU(2)$-Hilbert spaces - 
finite dimensional subspaces of the 
2-sphere square integrable functions  $L^2(\Om^2)$ -
 by starting from
$\SU(2)$-matrix elements
in the form of Fourier transformed mo\-men\-tum distributions 
serves as  a preparation for the $\SU(1,1)$-case below.
It is instructive to consider, side by side,
the re\-pre\-sen\-ta\-tions of the Euclidean group 
$\SU(2)\sx\R^3$ (rank 2),
relevant for  nonrelativistic scattering structures 
$\SO(3)\sx\R^3$ in 3-position,
and  of the spin (rotation) group $\SU(2)$ (rank 1).
 
The irreducible scalar matrix elements (spherical Bessel functions) 
with  Dirac measure for the momentum 2-sphere
\begin{eq}{rl}
\SU(2)\sx\R^3/\SU(2)\ni \rvec x\mape
{\sin \mu|\rvec x|\over \mu|\rvec x|}& 
=\int{d^3 p\over2\pi}~{1\over \mu}\de(\mu^2-\rvec p^2)e^{i\rvec p\rvec x}
\cr
&=\int{d^2\om\over4\pi}~e^{i\mu\rvec \om\rvec x}
=\int{d^2\om\over4\pi}~e^{-i\mu\rvec \om\rvec x},~\mu=P>0\cr
\end{eq}lead to those 
 with  derived Dirac distribution
\begin{eq}{rl}
\SU(2)\ni e^{i\rvec x}\mape\cos \mu|\rvec x|&= 
\int{d^3 p\over\pi}~\mu\de'(\mu^2-\rvec p^2)e^{i\rvec p\rvec x}
=\mu{\p\over\p\mu^2}\int{d^3 p\over\pi}~\de(\mu^2-\rvec p^2)
e^{i\rvec p\rvec x}\cr
&=\int{d^2\om\over4\pi}~(1+i\mu\rvec \om\rvec x)
e^{i\mu\rvec \om\rvec x}\cr 
&=(1+\rvec x{\p\over\p\rvec x})
\int{d^2\om\over4\pi}~
e^{i\mu\rvec \om\rvec x} ,~\mu=2J=1,2\dots\cr
\end{eq}The relevant integral sums over the 2-sphere mo\-men\-tum directions
 \begin{eq}{l}
\rvec \om
={\rvec p\over\sqrt{\rvec p^2}}=
{\scriptsize\pmatrix{
\cos\th& -ie^{i\phi}\sin\th\cr i e^{-i\phi}\sin\th&-\cos\th \cr }}
\in\Om^2,~~\rvec p=|\rvec p|\rvec\om\cr
\int {d^2\om\over 4\pi}=
\int_{-1}^1{d\cos\th\over2}\int_0^{2\pi}{d\phi\over2\pi},~~
\de({{\rvec \om}-{\rvec \om}{}'\over4\pi})=
\de({\cos\th-\cos\th'\over2})\de({\phi-\phi'\over2\pi})\cr
\end{eq}

For the neutral group element $\rvec x=0$,
there remains, for both underived and derived Dirac distribution,  
the  normalized mo\-men\-tum-sphere measure 
\begin{eq}{rl}
\int{d^3 p\over2\pi}~{1\over \mu}\de(\mu^2-\rvec p^2)
=\int{d^3 p\over\pi}~\mu\de'(\mu^2-\rvec p^2)=
\int{d^2\om\over4\pi}=1
\end{eq}The Hilbert space  relevant restriction   - in both cases -
is the Dirac distribution
on the 2-sphere.

The $\SU(2)\sx\R^3$-re\-pre\-sen\-ta\-tions are induced
by  translation re\-pre\-sen\-ta\-tions
with fixgroup $\SO(2)$, those for  $\SU(2)$ 
by $\SO(2)$-re\-pre\-sen\-ta\-tions
\begin{eq}{rrrr}
\hbox{for }\SU(2)\sx\R^3:&\R^3~\ni& \rvec x~~\mape& e^{iP\rvec\om\rvec x}~~\in\U(1)\cr
\hbox{for }\SU(2):&\SO(2)\ni& e^{i\rvec\om\rvec x\si_3}\mape& e^{i 2J\rvec\om\rvec x\si_3}
\in\SO(2)
\end{eq}

The   basis distribution
for the irreducible Hilbert space of the Euclidean group
contain scattering `states' 
 of mo\-men\-tum value $P$ in the direction $\rvec\om$ and
$\SO(2)$-polarization $\pm J=0,\pm {1\over 2},\pm 1,\dots$ with $\SO(2)\x\R^3$-action 
\begin{eq}{l}
(P^2,J)\in\irrep_+\SU(2)\sx\R^3:~~\left\{\begin{array}{l}
\{\rstate{P^2,J;\rvec \om,\ep}\mid 
\rvec \om\in\Om^2,\ep=\pm1\}\cr
\int{d^2\om\over4\pi}\rstate{P^2,J;\rvec \om,\ep}
\lstate{P^2,J;\rvec \om,\ep}\cong\id_{\C^2(\Om^2)}\cr
\rstate{P^2,J;\rvec \om,\ep}\stackrel{\SO(2)\x\R^3}\mape 
e^{ i\ep 2J\phi} e^{iP\rvec \om\rvec x}
\rstate{P^2,J;\rvec \om,\ep}
\cr
\end{array}\right.
\end{eq}$J=0$ has   `states' with  trivial polarization  
 $\{\rstate{P^2;\rvec \om}\}$. 
The `states' in the basis distributions for $\SU(2)$
with one invariant $J$ (spin) have 
 mo\-men\-tum direction $\rvec\om$ and
$\SO(2)$-eigenvalues $\ep J=\pm J$ 
\begin{eq}{l}
J\in\irrep\SU(2):~~
\left\{\begin{array}{l}
\{\rstate{J;\rvec \om,\ep}\mid \rvec \om\in\Om^2,\ep=\pm1\}\cr
\int{d^2\om\over4\pi}\rstate{J;\rvec \om,\ep}
\lstate{J;\rvec \om,\ep}\cong\id_{\C^2(\Om^2)}\cr

\rstate{J;\rvec \om,\ep}\stackrel{\SO(2)}\mape
e^{ \ep i2J\rvec\om\rvec x} \rstate{J;\rvec \om,\ep}\cr
\end{array}\right.
\end{eq}In addition to the
action of the subgroup
$\SO(2)$, the derived Dirac distribution 
leads also to matrix elements
with the action of the $\SO(2)$-Lie algebra
\begin{eq}{l}
\rstate{J;\rvec \om,\ep}\stackrel{\log\SO(2)}\mape
 \ep i 2J\rvec\om\rvec x \rstate{J;\rvec \om,\ep}\cr\end{eq}

The induced  action for the full rotation group $r\in\SU(2)$
\begin{eq}{l}
\rstate{P^2,J;\rvec \om,\ep}
\stackrel{\SU(2)}\mape o(r,\rvec\om)_\ep^{\ep'}
\rstate{P^2,J;\rvec \om,\ep'}
,~~\rstate{J;\rvec \om,\ep}
\stackrel{\SU(2)}\mape 
o(r,\rvec\om)_\ep^{\ep'}
\rstate{J;\rvec \om,\ep'}
\end{eq}comes as   Wigner axial rotation $o(r,\rvec\om)\in \SO(2)$ with
mo\-men\-tum-direction dependent parameters which is determined
by the rotation action on the 
representative $u(\phi,\th,0)\in \SU(2)/\SO(2)$ for the mo\-men\-tum direction orbit
\begin{eq}{rl}
r\in\SU(2):~ 
r\o u(\rvec\om)&=u(r\m\rvec\om)\o o(r,\rvec\om) \cr
\then o(r,\rvec\om)&=u(r\m\rvec\om)^\star\o r\o
u(\rvec\om)\in\SO(2)\cr
\hbox{with }u(\rvec\om)&=u(\phi,\th,0)
\in\SU(2)/\SO(2)
\end{eq}$r\m\rvec\om=r\o\rvec\om\o r^{-1}$
is the $\SU(2)$-rotated mo\-men\-tum direction.

This is in analogy to the familiar 
mo\-men\-tum dependent Wigner rotation
for an induced Lorentz transformation: There,
a  boost,  parametrized
by an energy-mo\-men\-tum vector $q\in\R^4$ on the hyperboloid $\cl Y^3$,
e.g. in the Weyl re\-pre\-sen\-ta\-tion 
\begin{eq}{l}
s({q\over m})=e^{\rvec\be\rvec\si} \in\SL(\C^2)/\SU(2),~~ 
{q^2\over m^2}=1,~~\sinh 2|\rvec \be|={|\rvec q|\over m}
\end{eq}is transformed by a Lorentz group action
$\SL(\C^2)\ni\la\sim \La\in\SO_0(1,3)$ into a boost for the transformed
mo\-men\-tum up to a Wigner rotation
\begin{eq}{l}
\la\o s({q\over m})=s(\La.{q\over m})\o r(\la,{q\over m}),~~ 
 r(\la,{q\over m})\in\SU(2)
\end{eq}

With orthogonal basis distributions
on the  mo\-men\-tum  2-sphere
\begin{eq}{rl}
\sprod{P^2,J;\rvec \om',\ep'}
{P^2,J;\rvec \om,\ep}&=\de^\ep_{-\ep'}\de({\rvec \om-\rvec \om{}'\over4\pi})\cr
\sprod{J;\rvec \om',\ep'}
{J;\rvec \om,\ep}&=\de^\ep_{-\ep'}\de({\rvec \om-\rvec \om{}'\over4\pi})
\end{eq}the  Hilbert vectors for $\SU(2)$ 
are  pairs of square integrable functions  
 $L^2(\Om^2)$ 
 \begin{eq}{rll}
\rstate{J;w}&=
\int {d^2\om\over 4\pi}
w(\rvec \om,\ep)\rstate{J;\rvec \om,\ep}&=
\int {d^3p\over 2\pi}~{1\over 2J}\de(\rvec p^2-4J^2)
w(\rvec p,\ep)\rstate{\rvec p,\ep}\cr
\sprod {J;w'}{J;w}&=
\int {d^2\om\over 4\pi}~
\ol{w'(\rvec \om,\ep)}w(\rvec \om,\ep)
&=\int {d^3p\over 2\pi}~
\ol{w'(\rvec p,\ep)}{1\over 2J}\de(\rvec p^2-4J^2)w(\rvec p,\ep)\cr
\end{eq}and analogously for the Euclidean group
where also position functions 
$w(\rvec p)=\int d^3x ~\tilde w(\rvec x)e^{i\rvec p\rvec x}$ 
can be used with a spherical Bessel function 
\begin{eq}{rl}
\sprod {P^2,J;w'}{P^2,J;w}&=\int {d^2\om\over 4\pi}~
\ol{w'(\rvec \om,\ep)}w(\rvec \om,\ep)\cr
&=\int {d^3p\over 2\pi}~
\ol{w'(\rvec p,\ep)}{1\over P}\de(\rvec p^2-P^2)w(\rvec p,\ep)\cr
&=\int {d^3xd^3x'\over 2\pi}~
\ol{\tilde w'(\rvec x',\ep)}{\sin P|\rvec x-\rvec x'|\over 
P|\rvec x-\rvec x'|}\tilde w(\rvec x,\ep)\cr
\end{eq}

Only for $\SU(2)$ with discrete invariant $\mu=2J$, a basis 
for the irreducible $(1+2J)$-dimensional Hilbert space
is given with the totally symmetrized $2J$-powers of the basis 
re\-pre\-sen\-ta\-tion  
$\SU(2)/\SO(2)\ni u(\phi,\th,0)$
 (spherical harmonics for integer $J$)
 - starting with the two functions in the  first and
 the three functions in the  middle 
column of the matrices for $J={1\over2}$ and $J=1$
\begin{eq}{l}
{\scriptsize\left(\begin{array}{c|c}
e^{i{\phi\over2}}\cos{\th\over2}
&ie^{i{\phi\over2}}\sin{\th\over2}\cr
ie^{-i{\phi\over2}}\sin{\th\over2}&
 e^{-i{\phi\over2}}\cos{\th\over2}\cr\end{array}\right)}\in\SU(2),~~
{\scriptsize\left(\begin{array}{c|c|c}
e^{i\phi}\cos^2{\th\over2}&
ie^{i\phi}{\sin \th\over\sqrt2}&
-e^{i\phi}\sin^2{\th\over2}\cr
i{\sin\th\over\sqrt2}&
\cos \th&
i{\sin\th\over\sqrt2}\cr
-e^{-i\phi}\sin^2{\th\over2}&
ie^{-i\phi}{\sin \th\over\sqrt2}&
e^{-i\phi}\cos^2{\th\over2}\cr\end{array}\right)}\in\SU(3)
\end{eq}The $\SU(1+2J)$-orthonormality of the 
columns  $\sprod{J;a'}{J;a}=\de_{-a'}^a$, which holds for any group element,
i.e. any $(\phi,\th)$,
has to be distinguished from 
Schur's orthogonality for the matrix elements
 which integrates over the full group. The 
 Euler angle pa\-ra\-me\-tri\-zed components in the
 vectors $\rstate{J;a}$ are Schur-orthogonal.

\subsection{Hilbert Spaces of  $\SU(1,1)$}

In the re\-pre\-sen\-ta\-tion matrix elements
for $(1,2)$-spacetime with an invariant $\mu>0$
\begin{eq}{l}
y\mape\int{d^3 q\over 2\pi}~{1\over \mu}\de(\mu^2-q^2)e^{iqy}
=\int{d^2\ty c\over4\pi}~e^{i\mu\ty c y}\cr
\hbox{ with }\left\{\begin{array}{l}
\ty c={q\over\sqrt{q^2}}
={\scriptsize\pmatrix{
\pm \cosh\psi&-e^{i\phi}\sinh\psi\cr
e^{-i\phi}\sinh\psi&\mp\cosh\psi\cr}}
\in\cl Y^2_\pm,~~q=|q|\ty c \cr
\int {d^2\ty c\over 4\pi} =\int {\sinh\psi~ d\psi\over2}\int_0^{2\pi}
{ d\phi\over 2\pi},~~\de({{\ty c}-{\ty c}{}'\over4\pi})=
\de({\cosh\psi-\cosh\psi'\over2})\de({\phi-\phi'\over2\pi})\cr

\end{array}\right.\cr
\cr\end{eq}\begin{eq}{l}

y\mape\int{d^3 q\over 2\pi}~{1\over \mu}\de(\mu^2+q^2)e^{iqy}
=\int{d^2\ty s\over4\pi}~e^{i\mu\ty s y} \cr
\hbox{ with }\left\{\begin{array}{l}
\ty s={q\over\sqrt{-q^2}}
={\scriptsize\pmatrix{
\sinh\psi&ie^{i\phi}\cosh\psi\cr
ie^{-i\phi}\cosh\psi&-\sinh\psi\cr}}
\in\cl Y^{(1,1)},~q=|q|\ty s \cr
\int {d^2\ty s\over 4\pi}
 =\int {\cosh\psi~ d\psi\over2} \int_0^{2\pi}{d\phi\over 2\pi},~~
\de({{\ty s}-{\ty s}{}'\over4\pi})
=\de({\sinh\psi-\sinh\psi'\over2})\de({\phi-\phi'\over2\pi})\cr
\end{array}\right.\cr
\end{eq}the unit vectors $(\ty c,\ty s)\in (\cl Y^2_\pm,\cl Y^{(1,1)})$ 
on the hyberboloids for the noncompact group $\SU(1,1)$
are the analogue to the sphere unit vectors $\rvec\om\in\Om^2$
for the compact group $\SU(2)$.

Energy-momentum functions
are expanded with basis distributions 
\begin{eq}{l}
\mu^2\in\irrep_+\SU(1,1):~
\left\{\begin{array}{l}
\{\rstate{\mu^2;q,\ep}\mid q\in \R^3,\ep=\pm1\}\cr
\int {d^3q\over(2\pi)^3}
\rstate{\mu^2;q,\ep}\lstate{\mu^2;q,\ep}
\cong\id_{\C^2(\R^3)}\cr
\rstate{\mu^2;w}=
\int d^3q~w(q,\ep)\rstate{\mu^2;q,\ep}\cr
\end{array}\right.\cr
\end{eq}The inducing Hilbert  
re\-pre\-sen\-ta\-tions of compact and noncompact  Cartan subgroup 
are powers with the invariant $\mu$ and $i\mu$
 \begin{eq}{rrll}
\SO(2)\ni& e^{i \ty c y\si_3}&\mape
(e^{ i\ty c y\si_3})^\mu &\in\SO(2)\cr
\SO_0(1,1)\ni& e^{ \ty s y\si_1}&\mape (e^{  \ty s y\si_1})^{i\mu}&\in\SO(2)
\end{eq}They act upon the basis distributions $\rstate{\mu^2;q,\ep}$
corresponding to the energy-momenta - either 
energylike $q=\ty c|q|$ or momentumlike $q=\ty s|q|$. 

The induced  actions for the full group $s\in\SU(1,1)$
\begin{eq}{l}
\rstate{\ty c|q|,\ep}
\stackrel{\SU(1,1)}\mape o(s,\ty c)_\ep^{\ep'}
\rstate{\ty c|q|,\ep'}
,~~
\rstate{\ty s|q|,\ep}
\stackrel{\SU(1,1)}\mape o(s,\ty s)_\ep^{\ep'}
\rstate{\ty s|q|,\ep'}
\end{eq}come in the form of  Wigner axial rotations 
$o(s,\ty c),~ o(s,\ty s)\in \SO(2)$ with
 parameters dependent on the hyperboloid points.
They can be constructed
 with the coset representatives above for the orbit pa\-ra\-me\-tri\-zations of the
 energy-mo\-men\-ta 
\begin{eq}{l}
s\in\SU(1,1)\then
 \left\{\begin{array}{rll}
\SO(2)\ni &v(s\m\ty c)^{-1}\o s\o
v(\ty c)&=o(s,\ty c)\cr
\SO_0(1,1)\ni& w(s\m\ty s)^{-1}\o s\o
w(\ty s)&\mape o(s,\ty s)\in\SO(2)\end{array}\right.\cr
\hbox{with }
 \left\{\begin{array}{rl}
v(\ty c)&=v(\phi,\psi,0)\in\SU(1,1)/\SO(2)\cr
w(\ty s)&=w(\phi,\psi,0)\in\SU(1,1)/\SO_0(1,1)
 \end{array}\right.
\end{eq}$s\m\ty c$ and  $s\m\ty s$ are the $\SU(1,1)$
transformed  directions on the hyperboloids.

For the  principal $\SU(1,1)$-re\-pre\-sen\-ta\-tions,
supported by a nontrivial
mo\-men\-tumlike  invariant $\mu=P$, 
 the  definition of - now - infinite dimensional  
irreducible Hilbert spaces with  square integrable functions 
 $L^2(\cl Y^{(1,1)})$ on the  mo\-men\-tumlike hyperboloid - 
  will be given in   analogy
to the $\SU(2)$-functions $L^2(\Om^2)$. 
The support of the Dirac distribution $\de(q^2+P^2)$ 
restricts the  basis distributions
from all energy-momenta $\R^3$ to the hyperboloid
$\cl Y^{(1,1)}$ 
\begin{eq}{rl}
P^2_\pm\in\irrep_+\SU(1,1):&
\left\{\begin{array}{l}
\{\rstate{P_\pm^2;\ty s,\ep}\mid \ty s\in \cl Y^{(1,1)},\ep=\pm1\}\cr
\int{d^2\ty c\over4\pi}
\rstate{P_\pm^2;\ty s,\ep}\lstate{P_\pm^2;\ty s,\ep}\cong\id_{\C^2(\cl Y^{(1,1)})}\cr
\rstate{P_\pm^2;\ty s,\ep}\stackrel{\SO_0(1,1)}\mape
e^{\pm \ep iP \ty s y} \rstate{P_\pm^2;\ty s,\ep}\cr
\end{array}\right.\cr
\end{eq}The derived Dirac distributions give  also  matrix elements
with the action of the $\SO_0(1,1)$-Lie algebra.
With the orthogonal scalar product distributions 
on the  hyperboloid
one obtains the product for the Hilbert vectors 
\begin{eq}{rl}
\sprod{P^2;\ty s',\ep'}
{P^2;\ty s,\ep}&=\de^\ep_{-\ep'}\de({{\ty s}-{\ty s}{}'\over4\pi})\cr
\rstate{P^2;w}&=\int {d^2\ty s\over4\pi}~w({\ty s},\ep)\rstate{P^2;{\ty s},\ep}
=\int {d^3q\over 2\pi}~
{1\over P}\de(q^2+P^2)w(q,\ep)\rstate{q,\ep}\cr
\sprod{P^2;w'}{P^2;w}
&=\int {d^2\ty s\over 4\pi}~
\ol{w'(\ty s,\ep)}w(\ty s,\ep)
=\int {d^3q\over 2\pi}~\ol{w'(q,\ep)}
{1\over P}\de(q^2+P^2)w(q,\ep)\cr
\end{eq}For $\SU(1,1)$, there is no finite dimensional
definite unitarity (no hyperbolic  harmonics) 
as seen also in the Euler `angle' group
pa\-ra\-me\-tri\-zation above.


The representations  for the discrete and supplementary series
with  energylike  invariant $\mu=n,m$
do not involve  a Dirac measure. The Hilbert spaces  are  
not characterized by square integrable functions. The Hilbert space functions
$\rstate{\mu^2,w}$  use  all  energy-mo\-men\-ta from  a  basis
distribution $\{\rstate{\mu^2;q,\ep}\mid
q\in\R^3\}$
with the inducing and  induced representations above.
The  energy-mo\-men\-tum dipoles in the 
representation matrix elements, valued 
in complex $(2\x2)$-matrices, i.e.  
in the $\SO_0(1,2)$ Lie algebra  
and - for the energy-momenta - in the Lie algebra forms 
\begin{eq}{rl}
\SU(1,1)\ni e^{iy}
&\mape
\left\{\begin{array}{ll}
\int{d^3q\over 2\pi^2}~{q\over [(q\mp io)^2-n^2]^2}e^{iqy}
&\hskip-3mm=-\vth(y^2)\vth(\pm y_0){iy\over|y|}\cos n|y|\cr
\int {d^3 q \over  \pi^2}{q\over (q_\ro P^2-m^2)^2}e^{iqy}
&\hskip-3mm=-\vth(y^2){iy\over|y|}\cos |my|\cr
\end{array}\right.\cr
 i y&={\scriptsize\pmatrix{iy_0&y_1-iy_2\cr
y_1+iy_2&-iy_0}}\in\log\SU(1,1)\cong\R^{1+2}\cr
\end{eq}lead to scalar matrix elements 
with  the invariant 
$y^2=\det iy={1\over2}\tr y\o y=
y_0^2-y_1^2-y_2^2$
\begin{eq}{l}
\det\Bigl[ \vth(y^2)\vth(\pm y_0){iy\over|y|}\cos \mu |y|\Bigr]
=\vth(y^2)\vth(\pm y_0)(\cos \mu |y|)^2
\end{eq}The corresponding 
Hilbert product distributions involve  
positive type functions 
  \begin{eq}{rl}
 \sprod{\mu^2;q',\ep'}{\mu^2;q,\ep}&=
 q^\ep_{-\ep'}\om_{\mu^2}(q^2)
 \de(q-q')\hbox{ with }\om_{\mu^2}(q^2)=\left\{\begin{array}{ll}
 {1 \over [(q\mp  io)^2-n^2]^2} \cr
  {1\over (q_\ro P^2-m^2)^2}
  \cr\end{array}\right.
\cr
\sprod{\mu^2;w'}{\mu^2;w}&
=\int d^3q 
~\ol{w'(q,\ep')} ~ q^\ep_{-\ep'}\om_{\mu^2}(q^2)~
 w(q,\ep )\cr
&=\int d^3q ~\om_{\mu^2}(q^2)
~\tr q\o (w\ox \ol{w'})(q)\cr
\end{eq}The Hilbert space function pairs $\{q\mape 
w(q,\ep)\mid \ep=\pm1\}$ are defined by 
$\sprod{\mu^2;w}{\mu^2;w}\ge0$
with the $(2\x2)$-matrix valued function
$q\mape (w\ox\ol w)(q)$, in analogy to the
energylike energy-momenta by ${1\over2}\tr q\o q=q^2\ge0$.

It remains to establish explicitly how the 
Hilbert spaces with ener\-gy-mo\-men\-tum functions,  of $L^2$-type
and with positive type function, 
are related to the Hilbert spaces as constructed with 
homogeneous functions
of one  variable $\xi\mape F(\xi)$ as mentioned above.

\newpage
\section{Appendix: $\C^3$-Lie Algebras}

For a real or complex finite dimensional Lie algebra $L$,
the classes with respect to the commutator ideal $[L,L]$, the
radical $R$ (maximal solvable ideal in $L$)  and the nilradical $N\sub R$ 
(maximal nilpotent ideal in $L$ and, also, in $R$)  are
\begin{eq}{c}
\begin{array}{|lcll|}\hline
\hbox{up to noncommutativity}&[L,L]&\then L/[L,L]  
&\hbox{abelian}\cr\hline
\hbox{up to solvability}&R&\then L/R\cong S\sub L  
&\hbox{semisimple}\cr\hline
\hbox{up to nilpotency}&N&\then L\cong L/N\rvec\pl N  
&\hbox{semidirect product}\cr\hline\end{array}
\end{eq}

Since there  are no semisimple Lie algebras with dimension 1 and 2,
the{ complex  3-di\-men\-sio\-nal Lie algebras} have radical 
$R\in\{\{0\},\C^3\}$, i.e. they 
are either simple or solvable.

They can be classified and constructed with the commutator ideal 
\begin{eq}{l}
L\cong\C^3\then
\p L=[L,L]\in\{\C^3,\C^2,\C,\{0\}\}
\end{eq}

The perfect case is the simple Lie algebra $A_1$
with a  basis $\{l^1,l^2,l^3\}$
for totally antisymmetric structure constants
\begin{eq}{l}
\C^3\cong \p L=A_1\hbox{ with }\left\{\begin{array}{ll}
[l^1,l^2]&=l^3\cr
[l^2,l^3]&=l^1\cr
[l^3,l^1]&=l^2\cr\end{array}\right.\cr
\end{eq}

The  ideals $\p L\cong\C^2$ belong to semidirect product
Lie algebras.  An abelian $\p L$ gives
a solvable, not nilpotent Lie algebra  with possible basis
\begin{eq}{l}
\begin{array}{l}
\p L=[L,\p L]=\C l^1\pl\C l^2,~~\p^2 L=\{0\}\cr
L=\C l^3\rvec\pl[\C l^1\pl\C l^2]\cr
\end{array}
\hbox{ with }\left\{\begin{array}{ll}
[l^1,l^2]&=0\cr
[l^2,l^3]&=l^1\cr
[l^3,l^1]&=l^2\cr\end{array}\right.
\cr
\end{eq}A semidirect $\C^2$-ideal  is not possible, since 
\begin{eq}{rl}
\p L=\C l^1\rvec\pl \C l^2:&\left\{\begin{array}{rl}
[l^1,l^2]&=l^2\cr
[l^2,l^3]&=\ga l^1+\de l^2\cr
[l^3,l^1]&=\al l^1+\be l^2\cr
\end{array}\right.\cr
\hbox{Jacobi identity: }&\left\{\begin{array}{rl}
0&=[l^1,[l^2,l^3]]+[l^2,[l^3,l^1]]+[l^3,[l^1,l^2]]\cr
&=\de l^2-\al l^2-\ga l^1-\de l^2\cr
\end{array}\right.\cr
\then (\al,\ga)=(0,0)
\then &\left.\begin{array}{rl}
[l^1,l^2]&=l^2\cr
[l^2,l^3]&=\de l^2\cr
[l^3,l^1]&=\be l^2\cr
\end{array}\right\}\then [L,L]=\C l^2 
\cr
\end{eq}

For $\p L\cong\C$ there is a basis with 
\begin{eq}{rl}
\p L=\C l^2:&
\left\{\begin{array}{rl}
[l^1,l^2]&=\be_1 l^2\cr
[l^2,l^3]&=\be_3 l^2\cr
[l^3,l^1]&=\al l^2\cr
\end{array}\right.\cr
\end{eq}$l^1$ and $l^3$ can be exchanged (1st and 2nd line). 
By renormalizations, a basis with three nontrivial brackets
leads to a basis with two nontrivial brackets,  and even to 
a basis with one nontrivial bracket
\begin{eq}{l}
\left\{\begin{array}{rl}
[l^1,l^2]&= l^2\cr
[l^2,l^3]&= l^2\cr
[l^3,l^1]&= l^2
\end{array}\right.
\iff
\left\{\begin{array}{rl}
[ l^1,l^2]&= l^2\cr
[\ul l^3,l^2]&= 0\cr
[\ul l^3, l^1]&= l^2\cr
\hbox{with } \ul l^3&= l^3+l^1\cr
\end{array}\right.
\iff
\left\{\begin{array}{rl}
[ l^1,l^2]&= l^2\cr
[\ul{\ul l}^3,l^2]&= 0\cr
[\ul{\ul l}^3, l^1]&= 0\cr
\hbox{with } \ul{\ul l}^3&=\ul l^3+l^2\cr
\end{array}\right.
\end{eq}which arises also from a basis with 
the following two nontrivial brackets
\begin{eq}{l}
\left\{\begin{array}{rl}
[l^1,l^2]&= l^2\cr
[l^3,l^2]&=l^2 \cr
[l^3,l^1]&=0\cr
\end{array}\right.\iff
\left\{\begin{array}{rl}
[l^1,l^2]&=l^2\cr
[\ul l^3,l^2]&=0 \cr
[\ul l^3, l^1]&=0\cr
\end{array}\right.,~~\ul l^3=l^3-l^1
\end{eq}This characterizes the decomposable  Lie algebra
$\C\pl[\C\rvec\pl\C]$.
 
Therefore, the only nondecomposable $\C^3$-Lie algebra 
with $\p L\cong\C$ is the nilcubic Heisenberg Lie algebra 
- in a basis  with two trivial brackets
\begin{eq}{l}
L=\C l^3\rvec\pl[\C l^1\pl \C l^2],~\p L=\C l^2,~[L,\p L]=\{0\}
\hbox{ with }\left\{\begin{array}{rl}
[l^1,l^2]&=0\cr
[l^2,l^3]&=0\cr
[l^3,l^1]&=l^2\cr
\end{array}\right.\cr
\end{eq}

\section{Appendix: Residual Distributions}

All residual re\-pre\-sen\-ta\-tions,
 considered  above, arise from   
the Fourier transformed generalized scalar functions\cite{GEL1} 
(where the $\Ga$-functions are defined with $\nu\in\R$) 
-  for linear  invariants
\begin{eq}{l}
\R:~~\int {d q \over 
2i \pi}{\Ga(1-\nu)\over
(q-io-m)^{1-\nu}}e^{iqx}
=\vth(x){ e^{imx}\over (ix)^\nu},~~m\in\R\cr
\end{eq}and  for quadratic invariants
in the scalar   distributions
for the definite orthogonal groups 
\begin{eq}{rl}
\begin{array}{c}
\O(d)\cr
d=2,3,\dots\cr
r=\sqrt{\rvec x^2}\cr
\end{array}
&\left\{\begin{array}{rl}
\int {d^d q\over\pi^{{d\over2}}} ~{\Ga({d\over2}-\nu)
\over (\rvec q^2)^{{d\over2}-\nu}}e^{i\rvec q\rvec x}&=
{\Ga(\nu)\over ({r^2\over4})^\nu}

\cr\cr
\int {d^dq\over\pi^{{d\over2}}} ~{\Ga({d\over2}-\nu)
\over (\rvec q^2+1)^{{d\over2}-\nu}}e^{i\rvec q\rvec x}&=
{2\cl K_{\nu}(r)\over ({r\over2})^\nu}\cr
\cr
\int {d^dq\over \pi^{{d\over2}}} ~{\Ga({d\over2}-\nu)
\over (\rvec q^2-io-1)^{{d\over2}-\nu}}e^{i\rvec q\rvec x}
&= {i\pi\cl H^{(1)}_{\nu}(r)\over ({r\over2})^\nu}

=-{\pi[\cl N_\nu-i\cl J_{\nu}](r)\over({r\over2})^\nu}

\cr
\end{array}\right.\cr
\end{eq}As seen in the power ${d\over2}$, there is a 
distinction between even and
odd dimensions $d$ - in parallel with the
Cartan series $D_R\cong\log\SO(\C^{2R})$ and
$B_R\cong\log\SO(\C^{1+2R})$.

The real-imaginary transition
relates to each other Macdonald and Hankel (with  Neumann and Bessel) functions
\begin{eq}{rl}
\xi\in\R:~~2\cl K_\nu(i\xi)&
=i\pi\cl H^{(2)}_\nu(\xi)=e^{i\pi\nu\over2}i\pi\cl H^{(1)}_\nu(-\xi)\cr
\pm i\cl H^{(1,2)}_{\nu}
&=-\cl N_{\nu}\pm i\cl J_{\nu},~~
\cl K_{-\nu}=\cl K_{\nu},~~
\cl H_{-\nu}^{(1)}=e^{i\pi\nu}\cl H_{\nu}^{(1)}

\end{eq}There are the 
special functions for $N=0,1,2,\dots$ - for halfinteger index
\begin{eq}{rl}
{({2\over\pi}\cl K_{N-{1\over2}},~
\cl J_{N-{1\over2}},~
 \cl N_{N-{1\over2}}
)( \xi )\over ({ \xi \over2})^{N-{1\over2}}}

&=(-{\p\over \p{ \xi ^2\over4}})^N 
{(e^{- \xi },~\cos \xi ,~\sin \xi )\over \sqrt\pi}\cr
{{2\over\pi}\cl K_{{1\over2}}(\xi)
\over ({ \xi \over2})^{-{1\over2}}}
={e^{- \xi }\over\sqrt\pi},~~
{[\cl N_{{1\over2}}\mp i\cl J_{{1\over2}}]( \xi )\over ({ \xi \over2})^{-{1\over2}}}
&=-{e^{\pm i\xi }\over\sqrt\pi},~~
{[\cl N_{-{1\over2}}\mp i\cl J_{-{1\over2}}]( \xi )\over ({ \xi \over2})^{-{1\over2}}}
=\mp i {e^{\pm i\xi }\over\sqrt\pi}\cr
\end{eq}and for integer index
\begin{eq}{rl}

{(\cl K_N,~\cl J_N,~\cl N_N)(\xi )\over 
({\xi \over2})^N}&=(-{\p\over\p{\xi ^2\over 4}})^N
(\cl K_0,~\cl J_0,~\cl N_0)(\xi )\cr
2\cl K_0(\xi )=\int d\psi~e^{-|\xi| \cosh\psi}&=
-{\SUM_{n=0}^\infty}{ ({\xi ^2\over4})^n\over (n!)^2}
[\log{\xi ^2\over4}-2\Ga'(1)-2\phi(n)]\cr
\lim_{\xi \to0}{\cl J_N(\xi )\over({\xi \over2})^N}&={1\over\Ga(1+N)},~~
\lim_{\xi \to0}({\xi \over2})^N\pi\cl N_N(\xi )=-\Ga(N)\cr
\end{eq}$\cl J_N$ has no singularities. The integer index functions
 give a quadratic dependence, e.g. in 
${\cl J_N(\xi)\over ({\xi\over2})^N}
=\cl E_N({\xi^2\over4})$.

By analytic continuation 
one obtains for indefinite orthogonal groups
\begin{eq}{l}
\begin{array}{c}
\O(n,m)\sx\R^d:\cr
n>0,~m>0\cr
d=2,3,\dots\cr
|x|=\sqrt{|x^2|}\cr\end{array}
~~\left\{\begin{array}{rl}
\int {d^d q \over 
 i^m\pi^{{d\over 2}}}{\Ga({d\over2}-\nu)\over
(q^2-io)^{{d\over2}-\nu}}e^{iqx}
& ={\Ga(\nu)\over \({x^2+io\over 4}\)^{\nu}}\cr
\cr
\int {d^dq\over i^m\pi^{{d\over2}}} ~
{\Ga({d\over2}-\nu)\over (q^2-io+1)^{{d\over2}-\nu}}&=
{\vth(x^2)
2\cl K_{\nu}(|x|) -\vth(-x^2)i\pi \cl H^{(2)}_{-\nu}(|x|)\over |{x\over2}|^\nu}
\cr
&\hfill-\de_{\nu}^N i\pi{\SUM_{k=1}^N}
{1\over (N-k)!}\de^{(k-1)}(-{x^2\over4})\cr
\end{array}\right.\cr
\end{eq}For integer $N=1,2,\dots$,
there arise $x^2=0$ supported Dirac distributions.

Orthogonally invariant  distributions 
are embedded in  hyperbolically invariant ones, e.g.
for $(1,s)$-spacetime with the general Lorentz groups
\begin{eq}{rl}
\begin{array}{ c}
\O(1,s)\cr
s=1,2,\dots\cr
N=1,2,\dots\cr
|x|=\sqrt{|x^2|}\cr
\end{array}&
\left\{\begin{array}{rl}
 \int {d^dq\over i^s
 \pi^{{d\over2}}} ~{\Ga({d\over2}-\nu)
\over (q^2-io)^{{d\over2}-\nu}}e^{iqx}&=

{\Ga(\nu)\over\({x^2+io\over4}\)^\nu}\cr
\cr
 \int {d^dq\over i^s
 \pi^{{d\over2}}} ~{\Ga({d\over2}-\nu)
\over (q^2-io+1)^{{d\over2}-\nu}}e^{iqx}&=

{\vth(x^2) 2\cl K_{\nu}(|x|)
-\vth(-x^2)\pi[\cl N_{-\nu}+i\cl J_{-\nu}](|x|) 
\over |{x\over2}|^\nu}\cr
&\hfill-\de_{\nu}^N i\pi{\SUM_{k=1}^N}
{1\over (N-k)!}\de^{(k-1)}(-{x^2\over4})\cr
\cr
\int {d^dq\over i^s\pi^{{d\over2}}} ~
{e^{i\nu\pi}\Ga({d\over2}-\nu)\over (q^2-io-1)^{{d\over2}-\nu}}e^{iqx}&=

{\vth(-x^2)2\cl K_{\nu}(|x|)
-\vth(x^2)\pi[\cl N_{-\nu}-i\cl J_{-\nu}]  (|x|)\over |{x\over2}|^\nu}
\cr
&\hfill+\de_{\nu}^N i\pi{\SUM_{k=1}^N}
{1\over (N-k)!}\de^{(k-1)}(-{x^2\over4})\cr
\end{array}\right.\cr
\end{eq}

For $\nu=-{1\over2}$ there are no singularities
 \begin{eq}{rl}
\O(1+s):
&\left\{\begin{array}{rl}
\int {d^{1+s} q\over\pi^{{2+s\over2}}} ~{\Ga({2+s\over2})
\over (\rvec q^2)^{{2+s\over2}}}e^{i\rvec q\rvec x}&=
-r\cr

\int {d^{1+s} q \over 
 \pi^{{2+s\over 2}}}{\Ga({2+s\over2})\over
(\rvec q^2+1)^{{2+s\over2}}}e^{i\rvec q\rvec x}
&
=e^{-r}\cr
\int {d^{1+s} q \over 
 \pi^{{2+s\over 2}}}{\Ga({2+s\over2})\over
(\rvec q^2-io-1)^{{2+s\over2}}}e^{i\rvec q\rvec x}
&
=ie^{ir}\cr
\end{array}\right.\cr
\end{eq}
\begin{eq}{rl}
\O(1,s):
&\left\{\begin{array}{rl}

 \int {d^{1+s} q \over 
i^s \pi^{{{1+s}\over 2}}}{\Ga({2+s\over2})\over
(q^2-io)^{{2+s\over2}}}e^{iqx}
 =&-|x|[\vth(x^2)+i\vth(-x^2)]\cr

 \int {d^{1+s} q \over 
i^s \pi^{{{1+s}\over 2}}}{\Ga({2+s\over2})\over
(q^2-io+ 1)^{{2+s\over2}}}e^{iqx}
 =&\vth ( x^2)e^{-|x|} +\vth (- x^2)e^{-i|x|}\cr
\int {d^{1+s} q \over 
i^{1+s} \pi^{{{1+s}\over 2}}}{\Ga({2+s\over2})\over
(q^2-io- 1)^{{2+s\over2}}}e^{iqx}
 =& \vth (- x^2)e^{-|x|}+\vth ( x^2)e^{i|x|}\cr

\end{array}\right.\cr

\end{eq}These distributions are
relevant for the re\-pre\-sen\-ta\-tions
of orthogonal groups in odd dimensions $\O(1+2R)$
and $\O(1,2R)$ with rank $R$ and  poles
of order $1+R=1,2,\dots$.  

For $\nu=0$ there is a logarithmic singularity
in $\cl K_0$ and $\cl N_0$
\begin{eq}{rl}
\O(1+s):
&\left\{\begin{array}{rl}

\int {d^{1+s} q \over 
 \pi^{{1+s\over 2}}}{\Ga({1+s\over2})\over
(\rvec q^2+1)^{{1+s\over2}}}e^{i\rvec q\rvec x}
&
 =2\cl K_{0}(r)\cr

\int {d^{1+s} q \over 
 \pi^{{1+s\over 2}}}{\Ga({1+s\over2})\over
(\rvec q^2-io-1)^{{1+s\over2}}}e^{i\rvec q\rvec x}
&=
-\pi [\cl N_0-i\cl J_0](r)\cr\end{array}\right.\cr
\end{eq}
\begin{eq}{rl}
\O(1,s):
&\left\{\begin{array}{rl}

\int {d^{1+s} q \over 
i^s \pi^{{{1+s}\over 2}}}{\Ga({{1+s}\over2})\over
(q^2-io+ 1)^{{{1+s}\over2}}}e^{iqx}
 =&\hskip-3mm
 \vth ( x^2)2\cl K_{0}(|x|)
-\vth (- x^2)\pi[\cl N_{0}+i\cl J_{0}](|x|)\cr
\int {d^{1+s} q \over 
i^s \pi^{{{1+s}\over 2}}}{\Ga({{1+s}\over2})\over
(q^2-io- 1)^{{{1+s}\over2}}}e^{iqx}
 =&\hskip-3mm
\vth (- x^2)2\cl K_{0}(|x|)-\vth ( x^2)\pi[\cl N_{0}-i\cl J_{0}](|x|)
\cr

\end{array}\right.\cr
\end{eq}These distributions are
relevant for the re\-pre\-sen\-ta\-tions
of orthogonal groups in even dimensions $\O(2R)$
and $\O(1,2R-1)$ with rank $R$  and  order $R=1,2,\dots$ poles.

The Fourier transformed simple poles are
used for  re\-pre\-sen\-ta\-tions of the Euclidean groups
$\SO(1+s)\sx\R^{1+s}$ and Poincar\'e groups $\SO_0(1,s)\sx\R^{1+s}$
\begin{eq}{rl}
\O(1+s):&\left\{\begin{array}{rl}

\int {d^{1+s}q\over\pi^{{1+s\over2}}} ~
{1\over\rvec q^2}e^{i\rvec q\rvec x}&=
{\Ga({s-1\over2})\over
({r\over2})^{s-1}}\cr
\cr

\int {d^{1+s}q\over\pi^{{1+s\over2}}} ~
{1\over\rvec q^2+1}e^{i\rvec q\rvec x}&=
{2\cl K_{{s-1\over2}}(r)\over
({r\over2})^{s-1\over2}}\cr

\cr\int {d^{1+s}q\over \pi^{{1+s\over2}}} ~
{1\over \rvec q^2-io-1}e^{i\rvec q\rvec x}&=
-{\pi[\cl N_{s-1\over2}-i\cl J_{s-1\over2}](r)\over ({r\over2})^{s-1\over2}}

\cr\end{array}\right.\cr
\end{eq}
\begin{eq}{rl}
\O(1,s):&\left\{\begin{array}{rl}
\int {d^{1+s}q\over i^s
 \pi^{{1+s\over2}}} ~{1
\over q^2-io}e^{iqx}&=
{\Ga({s-1\over2})\over\({x^2+io\over4}\)^{s-1\over2}}\cr
\cr
\int {d^{1+s}q\over i^s\pi^{{{1+s}\over2}}} ~{1\over q^2-io+1}e^{iqx}&=
{\vth(x^2) 2\cl K_{{s-1\over2}}(|x|)
-\vth(-x^2)\pi[\cl N_{-{s-1\over2} }+i\cl J_{-{s-1\over2}}](|x|)
\over |{x\over2}|^{s-1\over2}}\cr
&\hfill-\de_{s-1\over2}^N i\pi{\SUM_{k=1}^N}
{1\over (N-k)!}\de^{(k-1)}(-{x^2\over4})\cr
\cr
\int {d^{1+s}q\over i\pi^{{{1+s}\over2}}} ~
{1\over q^2-io-1}e^{iqx}&=
~{\vth(-x^2)2\cl K_{{s-1\over2}}(|x|)
-\vth(x^2)\pi[\cl N_{-{s-1\over2}}-i\cl J_{-{s-1\over2}}](|x|)
\over |{x\over2}|^{s-1\over2}}
\cr
&\hfill+\de_{s-1\over2}^N i\pi{\SUM_{k=1}^N}
{1\over (N-k)!}\de^{(k-1)}(-{x^2\over4})\cr
\end{array}\right.

\end{eq}The lightcone supported Dirac distributions arise for
even dimensional spacetime with nonflat position, i.e.
for $(1,s)=(1,3),(1,5),\dots$.

The one dimensional pole integrals
\begin{eq}{l}
\O(1)=\O(1,0):~~\left\{\begin{array}{rl}
 \int {d q \over  \pi}{1\over q^2+ 1}e^{iqx}
&=e^{-|x|}\cr
\int {d q \over  \pi}{1\over q^2-io- 1}e^{iqx}
 &=ie^{i|x|}\cr
\end{array}\right.
\end{eq}are spread  to 
odd dimensions starting with $1+s=3$ and a singularity at $|x|=0$
\begin{eq}{rl}
\O(3):&\left\{\begin{array}{rll}
\int {d^{3}q\over\pi^2} ~
{1\over\rvec q^2}e^{i\rvec q\rvec x}&=
{2\over r}\cr

\int {d^3 q \over   \pi^2}{1\over \rvec q^2+1}e^{i\rvec q\rvec x}
&=-{\p\over\p{r^2\over4}}e^{-r}&=2{e^{-r}\over r}\cr
\int {d^3 q \over 
 \pi^2}{1\over
\rvec q^2-io-1}e^{i\rvec q\rvec x}
&=-{\p\over\p{r^2\over4}}ie^{ir}&=2{e^{ir}\over r}\cr

\end{array}\right.

\cr\cr

\O(1,2):&\left\{\begin{array}{rl}

-\int {d^3q\over  \pi^2} ~{1
\over q^2-io}e^{iqx}&=
2{\vth(x^2)-i\vth(-x^2)\over|x|}
\cr
-\int {d^3 q \over  \pi^2}{1\over q^2-io+ 1}e^{iqx}
&=2{\vth(x^2)e^{-|x|}-\vth(-x^2)ie^{-i|x|}\over|x|}\cr
\int {d^3q \over i \pi^2}{1\over q^2-io- 1}e^{iqx}
&=2{\vth(-x^2)e^{-|x|}+\vth(x^2)ie^{i|x|}\over|x|}\cr

\end{array}\right.

\end{eq}The dipoles in three dimensions are  without singularity 
\begin{eq}{rl}
\O(3):&\left\{\begin{array}{rl}

\int {d^3 q \over   \pi^2}{1\over (\rvec q^2)^2}e^{i\rvec q\rvec x}
& =-r\cr
\int {d^3 q \over   \pi^2}{1\over (\rvec q^2+1)^2}e^{i\rvec q\rvec x}
& =e^{-r}\cr
\int {d^3 q \over 
 \pi^2}{1\over
(\rvec q^2-io-1)^2}e^{i\rvec q\rvec x}
&=ie^{ir}\cr
\end{array}\right.

\cr\cr

\O(1,2):&\left\{\begin{array}{rl}
-\int {d^3q\over  \pi^2} ~{1
\over (q^2-io)^2}e^{iqx}&=
|x|[\vth(x^2)-i\vth(-x^2)]
\cr

-\int {d^3 q \over  \pi^2}{1\over
(q^2-io+ 1)^2}e^{iqx}
&=\vth(x^2)e^{-|x|}+\vth(-x^2)e^{-i|x|}\cr

-\int {d^3 q \over  i\pi^2}{1\over (q^2-io- 1)^2}e^{iqx}
&=\vth(-x^2)e^{-|x|}+\vth(x^2)e^{i|x|}\cr
\end{array}\right.

\end{eq}

The 2-di\-men\-sio\-nal integrals integrate over the 
1-di\-men\-sio\-nal functions
\begin{eq}{rl}
\O(2):&\left\{\begin{array}{rll}
\int {d^2 q \over 
 \pi}{1\over
\rvec q^2+1}e^{i\rvec q\rvec x}
& =\int d\psi~e^{-r\cosh\psi}& =2\cl K_0(r)\cr
\int {d^2 q \over   \pi}{1\over
\rvec q^2-io-1}e^{i\rvec q\rvec x}
& =\int d\psi~e^{ir\cosh\psi}& =-\pi[\cl N_0-i\cl J_0](r)\cr

\end{array}\right.\cr\cr

\O(1,1):&\left\{\begin{array}{rl}

\int {d^{2} q \over i \pi}{1\over q^2-io+  1}e^{iqx}
 &=\vth ( x^2)2\cl K_0(|x|)-\vth (-x^2)\pi[\cl N_0+i\cl J_0](|x|) \cr
\int {d^{2} q \over i \pi}{1\over q^2-io- 1}e^{iqx}
 &=\vth (-  x^2)2\cl K_0(|x|)-\vth (x^2)\pi[\cl N_0-i\cl J_0](|x|) \cr
\end{array}\right.\cr\cr

\end{eq}They are spread to even dimensions - starting with $1+s=4$
\begin{eq}{rl}
\O(4):&\left\{\begin{array}{rll}

\int {d^{4}q\over\pi^2} ~
{1\over\rvec q^2}e^{i\rvec q\rvec x}&=
{4\over r^2}\cr

\int {d^4 q \over  \pi^2}{1\over\rvec q^2+1}e^{i\rvec q\rvec x}
&=-{\p\over\p {r^2\over4}}2\cl K_0(r)
& ={2\cl K_1(r) \over{r\over 2}}\cr
\int {d^4 q \over  \pi^2}{1\over \rvec q^2-io-1}e^{i\rvec q\rvec x}
&={\p\over\p {r^2\over4}}\pi
[\cl N_0-i\cl J_0](r)& =-{\pi[\cl N_1-i\cl J_1](r) \over{r\over 2}}\cr

\end{array}\right.\cr\cr

\O(1,3):&\left\{\begin{array}{rl}

-\int {d^4q\over 
 i\pi^2} ~{1
\over q^2-io}e^{iqx}&=
{4\over x^2_\ro P}-i\pi\de({x^2\over4})\cr

-\int {d^4 q \over  i\pi^2}{1\over q^2-io+ 1}e^{iqx}
 &= {\vth ( x^2)2\cl K_1(|x|)-\vth (- x^2)\pi[\cl N_{-1}+i\cl J_{-1}](|x|)
\over{|x|\over 2}}-
i\pi\de({x^2\over4})\cr

\int {d^4 q \over i \pi^2}{1\over q^2-io- 1}e^{iqx}
 &= {\vth (- x^2)2\cl K_1(|x|)-\vth ( x^2)\pi[\cl N_{-1}-i\cl J_{-1}](|x|)
\over{|x|\over 2}}\hfill+i\pi\de({x^2\over4})\cr
\end{array}\right.

\end{eq}Dipoles  for  four dimensions lead to maximally logarithmic  singularities.

\end{document}